\documentclass[11pt]{article}

\usepackage[in]{fullpage}

\usepackage{url}            
\usepackage{booktabs}       
\usepackage{amsfonts}       

\usepackage{subcaption}

\usepackage{mathrsfs}

\usepackage[table,xcdraw]{xcolor}
\definecolor{airforceblue}{rgb}{0.36, 0.54, 0.66}
\definecolor{amber}{rgb}{1.0, 0.75, 0.0}
\definecolor{burntorange}{rgb}{0.8, 0.33, 0.0}
\definecolor{dimgray}{rgb}{0.41, 0.41, 0.41}

\usepackage[pagebackref = true,linkcolor={red!80!black},citecolor={blue!70!black}, urlcolor={blue!70!black},colorlinks, breaklinks]{hyperref}
\usepackage{tikz}
\usetikzlibrary{fit}
\tikzset{%
	highlight/.style={rectangle,blend mode = multiply,draw=blue!90!black,thick,rounded corners = 0.3 mm,inner sep=0.5pt}
}

\usepackage{microtype}
\usepackage{graphicx}
\usepackage{booktabs} 

\usepackage{amsmath,amssymb,amsthm}
\usepackage[ruled,linesnumbered]{algorithm2e}
\usepackage{bbold}
\usepackage[sans]{dsfont}

\usepackage{pifont}

\usepackage{thmtools}
\usepackage{thm-restate}

\usepackage{multirow}

\usepackage{tabularx}

\usepackage{textcomp}

\usepackage[shortlabels]{enumitem}

\usepackage{thmtools}

\declaretheoremstyle[qed=$\square$,parent=section]{definitionwithend}

\declaretheorem[style=definitionwithend]{theorem}

\declaretheorem[style=definitionwithend]{definition}

\usepackage{tikz}
\usetikzlibrary{shadows}
\usetikzlibrary{arrows,positioning} 

\usepackage{contour}
\usepackage[normalem]{ulem}

\contourlength{0.8pt}

\numberwithin{equation}{section}

\usepackage{bm}

\usepackage{wrapfig}

\usepackage{authblk}
\newcommand{\PP}{\mathbb P}
\newcommand{\QQ}{\mathbb Q}
\newcommand{\EE}{\mathbb E}
\newcommand{\R}{\mathbb R}

\newcommand{\mc}{\mathcal}
\newcommand{\diff}{\mathrm{d}}
\newcommand{\eps}{\varepsilon}
\newcommand{\Let}{\triangleq}

\DeclareMathOperator*{\argmin}{argmin}

\def\O#1{\text{\ding{\the\numexpr#1+171}}}

\usepackage{threeparttable}
\allowdisplaybreaks[4] 

\def\keywords{\vspace{1.5em}
{\textbf{Keywords}:\,\relax%
}}

\usepackage{todonotes}
\usepackage{natbib}
\usepackage{tcolorbox}
\newtcbox{\alertinline}[1][red]
  {on line, arc = 0pt, outer arc = 0pt,
    colback = #1!20!white, colframe = #1!50!black,
    boxsep = 0pt, left = 1pt, right = 1pt, top = 2pt, bottom = 2pt,
    boxrule = 0pt, bottomrule = 1pt, toprule = 1pt}
\usepackage{newtxtext}
\newtcolorbox{textbox}[1]{
    sharp corners,
    boxsep=0mm,
    toptitle=2mm,
    lefttitle=0mm,
    colframe=black!3,
    colback=black!3,
    title={\rule[-2pt]{4.5pt}{10pt}\hspace*{1.5mm}#1},
    fonttitle=\bfseries\itshape\sffamily,
    coltitle=black,
    halign=flush left,
}
\usepackage[titletoc]{appendix}
\usepackage{lipsum}

\usepackage{setspace}

\begin{document}
\title{{
 Distributionally Robust Optimization and
Robust Statistics}}
\author[1]{Jose Blanchet\thanks{\href{mailto:jose.blanchet@stanford.edu}{jose.blanchet@stanford.edu}}}
\author[1]{Jiajin Li\thanks{\href{mailto:jiajinli@stanford.edu}{jiajinli@stanford.edu}}}
\author[1]{Sirui Lin\thanks{\href{mailto:siruilin@stanford.edu}{siruilin@stanford.edu}}}
\author[1]{Xuhui Zhang\thanks{\href{mailto:xzhang98@stanford.edu}{xzhang98@stanford.edu}}}
\affil[1]{Department of Management Science \& Engineering, Stanford University}
\date{\today}
   
\date{\today}

\maketitle

\vspace{-5mm}

\begin{abstract}
We review distributionally robust optimization (DRO), a principled approach for constructing statistical estimators that hedge against the impact of deviations in the expected loss between the training and deployment environments. Many well-known estimators in statistics and machine learning (e.g. AdaBoost, LASSO, ridge regression, dropout training, etc.) are distributionally robust in a precise sense. We hope that by discussing the DRO interpretation of well-known estimators, statisticians who may not be too familiar with DRO may find a way to access the DRO literature through the bridge between classical results and their DRO equivalent formulation. On the other hand, the topic of robustness in statistics has a rich tradition associated with removing the impact of contamination. Thus, another objective of this paper is to clarify the difference between DRO and classical statistical robustness. As we will see, these are two fundamentally different philosophies leading to completely different types of estimators. In DRO, the statistician hedges against an environment shift that occurs {\it{after}} the decision is made; thus DRO estimators tend to be pessimistic in an adversarial setting, leading to a min-max type formulation. In classical robust statistics, the statistician seeks to correct contamination that occurred {\it{before}} a decision is made; thus robust statistical estimators tend to be optimistic leading to a min-min type formulation. 

\end{abstract}

\begin{textbox}{Distributionally Robust Optimization (DRO)}
The task of DRO is to estimate a parameter that will perform well on an unseen population from samples generated from a given population, which may or may not be similar to the unseen population.
\end{textbox}
\begin{textbox}{Robust Statistics}
The task of robust statistics is to estimate a parameter that depends on a given population from samples that may be contaminated with outliers or errors. 
\end{textbox}
\newpage 
\tableofcontents
\newpage
\section{Introduction}
\begin{figure}[ht]
\centering

\tikzset{every picture/.style={line width=0.75pt}} 

\begin{tikzpicture}[x=0.75pt,y=0.75pt,yscale=-1,xscale=1]

\draw    (272,130.3) -- (339,130.3) ;
\draw [shift={(341,130.3)}, rotate = 180] [color={rgb, 255:red, 0; green, 0; blue, 0 }  ][line width=0.75]    (6.56,-1.97) .. controls (4.17,-0.84) and (1.99,-0.18) .. (0,0) .. controls (1.99,0.18) and (4.17,0.84) .. (6.56,1.97)   ;
\draw    (94,130.3) -- (161,130.3) ;
\draw [shift={(163,130.3)}, rotate = 180] [color={rgb, 255:red, 0; green, 0; blue, 0 }  ][line width=0.75]    (6.56,-1.97) .. controls (4.17,-0.84) and (1.99,-0.18) .. (0,0) .. controls (1.99,0.18) and (4.17,0.84) .. (6.56,1.97)   ;
\draw    (442,130.3) -- (509,130.3) ;
\draw [shift={(511,130.3)}, rotate = 180] [color={rgb, 255:red, 0; green, 0; blue, 0 }  ][line width=0.75]    (6.56,-1.97) .. controls (4.17,-0.84) and (1.99,-0.18) .. (0,0) .. controls (1.99,0.18) and (4.17,0.84) .. (6.56,1.97)   ;

\draw (153,53) node [anchor=north west][inner sep=0.75pt]   [align=left] {\begin{minipage}[lt]{74.18pt}\setlength\topsep{0pt}
\begin{center}
Contaminated \\data-generating\\distribution
\end{center}

\end{minipage}};
\draw (349,62) node [anchor=north west][inner sep=0.75pt]   [align=left] {\begin{minipage}[lt]{55.45pt}\setlength\topsep{0pt}
\begin{center}
Data-driven\\model
\end{center}

\end{minipage}};
\draw (512,62) node [anchor=north west][inner sep=0.75pt]   [align=left] {\begin{minipage}[lt]{67.35pt}\setlength\topsep{0pt}
\begin{center}
Out-of-sample\\environment
\end{center}

\end{minipage}};
\draw (48,121) node [anchor=north west][inner sep=0.75pt]    {$\mathbb{P}_{\star }$};
\draw (192,121) node [anchor=north west][inner sep=0.75pt]    {$\ \overline{\mathbb{P}}$};
\draw (372,121) node [anchor=north west][inner sep=0.75pt]    {$\ \ \hat{\mathbb{P}}_{n} \ $};
\draw (543,121) node [anchor=north west][inner sep=0.75pt]    {$\ \ \widetilde{\mathbb{P}}$};
\draw (182,184) node [anchor=north west][inner sep=0.75pt]    {$\begin{aligned}
 & \mathbb{\widetilde{\mathbb{P}} \ =P}_{\star } =\overline{\mathbb{P}} :\ \text{Conventional Assumption}\\
 & \widetilde{\mathbb{P}} \neq \hat{\mathbb{P}}_{n} :\ \text{Overfitting}\\
 & \mathbb{\widetilde{\mathbb{P}} \neq P}_{\star } =\overline{\mathbb{P}} :\ \text{Distributional Shift}\\
 & \mathbb{\widetilde{\mathbb{P}} =P}_{\star } \neq \overline{\mathbb{P}} :\ \text{Data Contamination}
\end{aligned}$};

\end{tikzpicture}
\caption{Data-Driven Decision Making Cycle} 
\label{fig:decisioncycle}
\end{figure} 




In the conventional data-driven decision-making cycle shown in Figure~\ref{fig:decisioncycle}, we typically observe $n$ i.i.d.~samples generated from the unknown data-generating distribution $\PP_{\star}$, and make a decision (e.g., parameter estimation) based on a model, $\hat{\PP}_n$, built from these samples (such a model could be parametric or non-parametric). These decisions are then deployed in the out-of-sample environment $\tilde{\PP}$, which may or may not follow the distribution $\PP_{\star}$.
During this cycle, several factors can contribute to suboptimal decision-making. 

 \textrm{(i)} \textit{Potential Overfitting} ($\widetilde{\PP} \neq \hat{\PP}_n $): when the sample size $n$ is not large enough, the model learned from the samples may achieve good in-sample performance but fail to generalize its predictive power to the out-of-sample environment, which is commonly termed as overfitting in statistical and machine learning tasks. Specifically, when referring to the out-of-sample environment in this scenario, people usually focus on the data-generating distribution ${\PP}_\star$. 

  \textrm{(ii)} \textit{Distributional shift} ($\widetilde{\PP} \neq {\PP}_\star = \overline{\PP}$): in many real-world scenarios, the out-of-sample environment $\tilde{\PP}$ may deviate from the data-generating distribution ${\PP}_\star$. This discrepancy, known as distributional shift, can arise in various circumstances. For instance, in adversarial deployment settings, malicious actors can intentionally manipulate the data distribution to undermine the performance of trained models. Additionally, in transfer learning settings, models may be expected to effectively generalize to target datasets that differ slightly from the source datasets used for training.

 \textrm{(iii)} \textit{Data Contamination} ($\widetilde{\PP} = {\PP}_\star \neq \overline{\PP}$): 
    many real data sets contain outliers or have measurement error throughout the steps of data generation and collection. Thus, the observed samples are actually generated by a contaminated distribution $\overline{\PP}$, posing challenges to the inference of the underlying uncontaminated distribution $\PP_\star$.



The first two cases, (i) and (ii), arise from errors that occur in the post-decision stage, where the trained model or decision rule is applied to the out-of-sample data. The third case, (iii), differs from these two in that the error occurs in the pre-decision stage, during the steps of data generation and collection. Now we discuss two principled approaches to deal with cases (i)-(iii). 
 
Distributionally robust optimization (DRO) is a data-driven decision-making framework that is designed to minimize the potential discrepancies between the in-sample expected loss and the out-of-sample expected loss. In particular, the goal of DRO is to address cases (i) and (ii). DRO takes an adversarial formulation aiming to minimize the expected loss of a model (e.g. the squared loss in linear regression) incurred by a parameter selection (e.g. the regression parameter), uniformly across a set of possible data distributions. The set of possible data distributions is characterized by a family of models that describe deviations from the training data distribution. To establish a concrete mathematical formulation, we begin by examining a generic stochastic optimization problem. 
Here, we assume that $\xi$ is a random vector in space $\Xi$ (e.g., $\R^d$) that follows the distribution $\PP_{\star}$. The set of feasible model parameters is denoted $\Theta$ (assumed to be finite-dimensional to simplify). Given a realization $\xi$ and a model parameter $\theta \in \Theta$ the corresponding loss is $\ell(\theta, \xi)$. A standard expected loss minimization decision rule is obtained by solving
\begin{align}\label{eq:ERM}
    \min_{\theta \in \Theta} ~\EE_{\PP_\star}[\ell(\theta, \xi)] = \int_{\Xi} \ell(\theta, \xi)~\diff \PP_\star (\xi).
\end{align}
Since $\PP_{\star}$ is generally unknown, to approximate the objective function in \eqref{eq:ERM}, we often i.i.d. samples $\xi_1, \dots, \xi_n$ each following distribution $\PP_\star$ and consider the empirical risk minimization counterpart,  
\begin{align}\label{eq:erm}
    \min_{\theta \in \Theta} ~\EE_{\hat \PP_n}[\ell(\theta, \xi)] = \frac{1}{n} \sum_{i=1}^{n} \ell(\theta, \xi_i),
\end{align}
where $\hat \PP_n$ denotes the empirical measure $\frac{1}{n}\sum_{i=1}^{n}\delta_{\xi_i}$ and $\delta_{\xi}$ is the Dirac measure centered at $\xi$.
To guarantee a good performance in the sense of finding a bound on the optimal expected population loss with high probability when deployed out-of-sample, the DRO framework introduces an uncertainty set $\mc B(\hat \PP_n)$ to capture variations between the in-sample distribution $\hat{\PP}_n$ and the out-of-sample distribution (including environment shifts). Then, the DRO formulation minimizes the worst-case loss within this uncertainty set, i.e., 
\begin{align}\label{eq:DRO}
    \min_{\theta \in \Theta} \sup_{\QQ \in \mc B(\hat \PP_n)} \EE_{\QQ}[\ell(\theta, \xi)],
\end{align}
where $\EE_Q$ denotes the expectation operator assuming that $\xi$ follows distribution $\QQ$. In Section~\ref{sec:DROsec}, we provide a comprehensive discussion of DRO and its effectiveness in addressing cases (i) and (ii). We examine its theoretical foundations and present evidence from diverse applications in statistical and machine learning communities. In particular, we discuss how DRO recovers a wide range of successful estimators and regularization methods, which  are often applied in statistical inference. We also discuss novel uses of DRO strategies that combine historical data with additional domain-knowledge information.

Case (iii) is thoroughly examined in the field of robust statistics, which seeks to address the challenges in inference posed by this case. Our goal here is to discuss a novel perspective that motivates the types of estimators that are obtained in robust statistics and explain their qualitative differences relative to DRO-based estimators. Traditional robust statistics techniques aim to estimate a parameter that depends on an unknown population from its contaminated samples. Formally, given samples $\xi_1, \dots, \xi_n$ in a metric space $\Xi$ that are i.i.d.~generated from some distribution $\bar \PP$ that may be a contaminated version of $\PP_{\star}$, we aim to learn a mapping $\hat \theta: \mc P(\Xi) \rightarrow \Theta$ that maps the empirical distribution $\hat \PP_n$ of the samples to an estimator $\hat \theta(\hat \PP_n)$ of the underlying parameter under $\PP_{\star}$. The contamination model can be formally represented as $\bar\PP\in\mathcal{A}(\PP_\star)$, where $\mathcal{A}(\PP_\star)$ is a set of possible contaminated data generating distributions that contains $\PP_\star$. Therefore, for a loss function $\ell: \Theta \times \Xi \rightarrow \R$, the out-of-sample risk given the robust learning procedure $\hat \theta(\hat \PP_n)$ is thus 
\[
\EE_{\PP_\star}[\ell(\hat\theta(\hat\PP_n), \xi)].
\]
The ``adversary'' that injects contamination is attempting to maximize this out-of-sample risk, while the decision-maker's goal is to minimize it. It is thus natural to introduce a max-min game to formalize this process as
\[\sup_{\bar\PP\in\mathcal{A}(\PP_\star)}\inf_{\hat\theta(\cdot)\in\Psi}\EE_{\bar\PP}\left[\EE_{\PP_\star}[\ell(\hat\theta(\hat\PP_n), \xi)]\right].\]
Here, we use $\bar \PP$ in short for the product measure $\bar \PP^{\otimes n}$. $\Psi$ is a class of policies that the decision-maker can employ, where each policy is a mapping $\hat \theta: \mc P(\Xi) \rightarrow \Theta$. Note that in this formulation, the inner randomness comes from $\xi$ that follows the distribution $\PP_\star$, and the outer randomness lies in the empirical measure $\hat \PP_n$, where each sample i.i.d. follows the distribution $\bar \PP$ . Given that the statistician knows that the data has been contaminated, a natural policy class to consider involves rectifying/correcting the contamination, and, for this, we introduce a rectification set $\mc R(\hat \PP_n)$ which models a set of possible pre-contamination distributions based on the knowledge of the empirical measure $\hat \PP_n$. The rectification/decontamination approach naturally induces the following min-min strategy to address the error in case (iii), thus 
\begin{align*}
    \hat\theta(\hat\PP_n) = \mathop{\arg\inf}_{\theta \in \Theta}\min_{\QQ \in\mc R (\hat {\PP}_n)}\EE_{\QQ}\left[\ell(\theta, \xi)\right].
\end{align*}
In Section~\ref{sec:robuststat}, we furnish a detailed overview of robust statistics, making explicit connections with the min-min approach that we present here while drawing comparisons with the DRO framework.

It is important to highlight the distinctive features of the DRO and the robust statistics formulations induced by varying the order of decision-making relative to the distributional mismatch in the three cases. In both cases (i) and (ii), the distributional mismatch occurs in the post-decision stage. Consequently, the DRO approach employs a min-max game strategy to control the worst-case loss over potential post-decision distributional shifts. In contrast, for case (iii), the robust estimator acts after the pre-decision distributional contamination materializes. Thus the approach of robust statistics can be motivated as being closer to a max-min game against nature. As a consequence, in robust statistics, the adversary moves first, and therefore the statistician can be more optimistic that they can rectify the contamination applied by the nature thus motivating the min-min strategy suggested above.

As we present our discussion, we will often summarize key results in the form of theorems which are stated in a summarized form for ease of exposition. We refer the reader to the references for precise assumptions and proofs. 

The DRO literature is rapidly growing, so it is virtually impossible to cover every new application in this review. However, in the conclusion section, we will briefly discuss trending topics on DRO in areas such as dynamic decision-making problems \citep{xu2010distributionally, osogami2012robustness, lim2013reinforcement, zhou2021finite, backhoff2022estimating,si2023distributionally,wang2023foundation} and causal inference~\citep{bertsimas2022distributionally, rothenhausler2023distributionally,bennett2023minimax,  duchi2023distributionally}.


\vspace{2mm}
\noindent \textbf{Notation}. To summarize the notation that we use, it is useful to keep in mind Figure~\ref{fig:decisioncycle}. We use $O(\delta)$ to denote a quantity that is bounded by $\delta \times$ some constant as $\delta$ goes to zero; we use $\mathbb{I}_{A}$ to denote the indicator function of the set $A$; we use $\delta_{\xi}$ to denote the Dirac measure at $\xi$ and let $\hat \PP_n \Let \frac{1}{n} \sum_{i=1}^n \delta_{\xi_i}$ be the empirical measure constructed from observed samples $\{\xi_1,\dots,\xi_n\}$; we use $\PP_{\star}$ to denote the underlying uncontaminated distribution, $\overline{\PP}$ to denote the (possibly contaminated) data-generating distribution, $\widetilde{\PP}$ to denote the out-of-sample distribution; in Section~\ref{sec:robuststat}, we also denote by $\bar\PP_n$ the contaminated version of $\hat\PP_n$; $\EE_{\PP}$ is the expectation over the probability distribution $\PP$; for a joint distribution $\pi$ for $(\xi, \eta)$, $\pi_{\xi}$ denotes the marginal distribution of $\xi$; $\overset{d}{\rightarrow}$ denotes the convergence in distribution; $\mc L^2(\mc D)$ denotes the $L^2$-integrable functions defined on domain $\mc D \subset \R^d$ under the Lebesgue measure on the $d$-dimensional Euclidean space; $\R_{+}$ denotes the space of non-negative real numbers; $\mathds{D}$ denotes discrepancies between probability models;   $\mc B$ denotes the uncertainty set of distributions;  $\mc R$ denotes the rectification set of distributions; $\mc A$ denotes the class of possible contaminated data generating distributions; $\Psi$ (resp., $\Phi$) denotes the class of policies that the decision maker can employ in the setting of robust statistics (resp., DRO).

\section{Distributionally Robust Optimization}\label{sec:DROsec}

The DRO framework provides a principled approach to understand and analyze various regularization methods from a probabilistic perspective. In this section, we will present a comprehensive overview of how the DRO problem \eqref{eq:DRO} connects with some well-known regularization methods commonly used in statistical and machine learning tasks. Throughout this overview, we also discuss how the DRO framework helps extend existing methods for these tasks by modeling distributional uncertainty in an interpretable adversarial way. Following this discussion, we review some of the statistical guarantees obtained by DRO-based estimators and the tools for statistical inference that are induced by the DRO framework, including new statistical objects such as associated worst-case distributions often corresponding to a Nash equilibrium.


Note that, in this section, $\PP_{\star}$ denotes the (uncontaminated) data-generating distribution because the error is assumed to occur in the post-decision stages.

\subsection{DRO Formulations and Related Statistical Tasks}\label{sec:drostattasks}

To be able to control the model’s conservativeness, 
most of existing literature define the uncertainty set $\mc B$ as a neighborhood ball that contains the nominal distribution $\hat \PP_n$ (e.g., empirical distribution). In this subsection, we review various probabilistic characterizations used in the DRO literature to construct the uncertainty set $\mc B$. Then, we connect the resulting DRO formulation with various statistical and machine learning tasks, which are summarized in Table~\ref{tab:stattasks}.
\begin{table}
\tabcolsep=0pt
\caption{Statistical tasks related to  DRO. (OT is short for optimal transport, $\phi$-div is short for $\phi$-divergence, CR is short for confidence region, and MMD is short for maximum mean discrepancy.)} 
\begin{tabular*}{\columnwidth}{@{\extracolsep{\fill}}l@{\hskip -0.05in}c@{\hskip -0.05in}c@{}} 
\hline
\rule{0pt}{2.0\normalbaselineskip}
Statistical tasks & 
\shortstack{Uncertainty  \\ construction} &  Reference\\
\hline
Norm regularization & OT &  
\citet{blanchet2019robust, li2022tikhonov, gao2022wasserstein}\\
\shortstack{Variance \\ regularization} & $\phi$-div & \citet{lam2016robust, lam2018sensitivity, duchi2021statistics} \\
Adaptive boosting &  $\phi$-div  & \citet{blanchet2019distributionally}\\
Domain adaptation & OT/$\phi$-div & \citet{taskesen2021sequential, zhang2022class}\\
Group regularization & OT/$\phi$-div & \citet{blanchet2017distributionally,hu2018does, sagawa2019distributionally}\\
Bayesian estimation & OT/$\phi$-div & \citet{zhang2022wasserstein, nguyen2023bridging, lotidis2023wasserstein}\\
CR construction & OT/$\phi$-div &
\citet{duchi2021statistics, he2021higher, blanchet2022confidence}\\
Hypothesis testing & OT/$\phi$-div & \citet{gul2017minimax, gao2018robust, sun2021data}\\
\rule{0pt}{2.0\normalbaselineskip}\shortstack{Dropout \\ regularization} &\shortstack{Multiplication  \\ auxiliary} & \citet{blanchet2020machine} \\
\hline
\end{tabular*}
\label{tab:stattasks}
\end{table}

A natural approach for modeling the distributional uncertainty set in DRO is given by moment constraints; in fact, this is one of the earlier approaches followed by~\citet{scarf1958minmax}. For example, a distributionally robust moment constraint problem involving means and variances was introduced in~\citet{delage2010distributionally} and it is formulated via 
\begin{align*}
    \min_{\theta \in \Theta} \sup_{\QQ \in \mc B} \EE_{\QQ}[\ell(\theta, \xi)],
\end{align*}
where the set $\mc B$ is defined as follows. Specify $\delta_1, \delta_2$, non-negative constants and estimate nominal mean and covariance matrix $(\hat \mu, \hat \Sigma)$ from the empirical distribution $\hat {\PP}_n$. Then,
\begin{align*}
    \mc B = 
    \left\{
    \QQ:
    (\EE_{\QQ}[\xi] - \hat \mu)^\top {\hat \Sigma}^{-1} (\EE_{\QQ}[\xi] - \hat \mu) \leq \delta_1,\ 
     \EE_{\QQ}[(\xi - \hat \mu)^\top (\xi - \hat \mu)] \preceq \delta_2 \hat \Sigma
    \right\}.
\end{align*}
This formulation has significant computational advantages discussed in \citet{delage2010distributionally} because it often can be formulated in terms of semi-definite programming. However, the problem with this formulation from a statistical standpoint is that $\mc B$ contains excessive distributions that are not in the local neighborhood of $\hat \PP_n$; even when $\delta_1, \delta_2$ are close to zero. As $n$ grows to infinity, under mild assumptions (certainly under i.i.d. assumptions) the nominal distribution $\hat \PP_n$ converges weakly with probability one to the data-generating distribution $\PP_{\star}$. Nevertheless, the uncertainty set contains distributions that may be far from $\PP_{\star}$ potentially deteriorating the performance of the DRO-based estimator in the sense of being over-conservative (except in cases in which the optimal parameter choice only depends on means and variances). 

Alternatively, most of the existing literature adopts uncertainty sets that are induced by probability metrics/discrepancies, that is, 
\begin{align*}
    {\mc B}_{\delta}(\hat \PP_n) = \left\{\QQ: \mathds{D}(\QQ, \hat \PP_n) \leq\delta\right\}.
\end{align*}
This ball of radius $\delta \ge0$ is defined with respect to a discrepancy measure $\mathds D$ on the probability distribution space $\mc P(\Xi)$. In the remainder of this section, we will focus on typical uncertainty sets $\mc B_\delta(\hat{\PP}_n)$ that are induced by probability metrics/discrepancies. This focus aims to serve our discussion on the connection of DRO with statistical and machine learning tasks. For a more in-depth discussion of general DRO formulations, interested readers are referred to~\citet{rahimian2019distributionally}.


\subsubsection{{$\phi$-Divergence-Based DRO}} The $\phi$-divergence approach corresponds to methods that penalize deviations from a baseline model in terms of the likelihood ratio; see, for example, \citet{csiszar1975divergence,ruszczynski2006optimization, rockafellardistributional}.
\begin{definition}[$\phi$-divergence]\label{def:phidiv}
    Assume that 
 $\phi: \R_+ \rightarrow (-\infty, +\infty]$ is a convex function with $\phi(0) = \lim_{t\rightarrow 0^+} \phi(t)$, then the $\phi$-divergence between $\QQ$ and $\hat \PP_n$ is
    \begin{align*}
        \mathds{D}_{\phi}(\QQ, \hat \PP_n) = 
        \left\{
        \begin{array}{ll}
             \int_{\Xi} \phi\left(\frac{\diff \QQ}{\diff \hat \PP_n}\right) \diff \hat \PP_n(\xi) &\quad \QQ \ll \hat \PP_n  \\
             +\infty &\quad \text{otherwise}, 
        \end{array}
        \right.
    \end{align*}
    where $\frac{\diff \QQ}{\diff \hat \PP_n}$ is the likelihood ratio between $\QQ$ and $\hat \PP_n$, and $\QQ \ll \hat \PP_n$ indicates that $\QQ$ is absolutely continuous with respect to $\hat \PP_n$.
\end{definition}

In the context of data-driven estimation using empirical measures, the approach is closely related to the empirical likelihood~\citep{owen2001empirical} method. In the data-driven setting, the works of \citet{hu2013kullback, lam2016robust, lam2018sensitivity, duchi2018variance, duchi2021learning} provide example showing the utilization of  
$\phi$-divergence to define the uncertainty set in various statistical tasks. 

The intuition using $\phi$-divergence is that the adversary can re-weight the relative importance of each sample with a budget constraint. So, the adversary systematically explores how re-weighting can potentially impact the performance of an estimator as measured by a given expected loss.

In~\citet{duchi2021learning}, the authors argue that this choice of uncertainty set (with a well-chosen $\phi$) can be used to hedge against the potentially low performance of statistical loss in minority subpopulations. Intuitively, if a minority population is severely affected by a decision choice, the adversary will exploit this by increasing the importance of this minority population, thus encouraging the decision maker to make a more equitable decision rule.

When the function $\phi(\cdot) \geq 0$ is twice differentiable and locally strongly convex around 1 (in particular, $\phi(1)=0, \phi'(1)=0, \phi''(1) > 0$), then $\phi$-divergence-based DRO approach is asymptotically equivalent to variance regularization. This equivalence is formally established by the following theorem: 

\begin{theorem}[Variance regularization {\citep[Theorem 3.1]{lam2016robust}, \citep[Theorem 1]{duchi2018variance}}]\label{thm:var-reg}
    Suppose that $\phi$ is twice differentiable around 1, and for simplicity assume that loss function $\ell$ is bounded, then we have
    \begin{align*}
    \min_{\theta \in \Theta} \sup_{\QQ \in {\mc B}_{\delta}(\hat \PP_n)} \EE_{\QQ}[\ell(\theta, \xi)] 
    = \min_{\theta  \in \Theta} \EE_{\hat{\PP}_n}[\ell(\theta, \xi)] + \sqrt{\frac{2 \delta}{\phi^{''}(1)} \textrm{Var}_{\hat \PP_n}(\ell(\theta, \xi))} + O(\delta),
\end{align*}
where ${\mc B}_{\delta}(\hat \PP_n) = \{\QQ: \mathds{D}_{\phi}(\QQ, \hat \PP_n) \leq \delta\}$,
$\textrm{Var}_{\hat \PP_n}(\ell(\theta, \xi)) = \EE_{\hat \PP_n}[\ell(\theta, \xi)^2] - (\EE_{\hat \PP_n}[\ell(\theta, \xi)])^2$ is the empirical variance of $\ell(\theta, \xi)$ under the empirical distribution $\hat \PP_n$.
\end{theorem}

This result can be intuitively expected with a back-of-the-envelope calculation as follows. Suppose that the center of the distributional uncertainty set is $\PP$ (this is more general than the empirical measure $\hat \PP_n$ in the data-driven setting illustrated in the theorem) and assume that $Z$ is the likelihood ratio of $\frac{\diff \QQ}{\diff \PP}$, such that $\EE_{\QQ}[\ell(\theta, \xi)]= 
\EE_{\PP}[\ell(\theta, \xi)\cdot Z]$ and $\mathds{D}_{\phi}(\QQ, \PP) = \EE_{\PP}[\phi(Z)]$. 

Under rather mild assumptions, the constraint $\mathds{D}_{\phi}(\QQ, \PP)=\EE_{\PP}[\phi(Z)]\leq\delta$ will be active when $\delta$ is small. Thus, with $\phi(1)=0, \phi'(1) = 0, \phi''(1) > 0$, upon the Taylor expansion on $\phi(\cdot)$ around $1$, the constraint will basically correspond to $\phi''(1)\EE_{\PP}[(Z-1)^2]/2 = \delta$. We thus write $Z = 1 + \delta^{1/2} \times \Delta$, where $\EE_{\PP}[|\Delta|^2] = 2 \delta / \phi''(1)$, and $\EE_{\PP}[\Delta] = 0$ to preserve that $Z$ is a likelihood ratio. We can ignore the positivity constraint on $Z$ since $\delta$ will go to zero. As a result, the adversary is essentially maximizing $\EE_{\QQ}[\ell(\theta, \xi)] = \EE_{\PP}[\ell(\theta, \xi)\cdot Z] = \EE_{\PP}[\ell(\theta, \xi)(1+\delta^{1/2} \Delta)]$, subject to the indicated constraints on $\Delta$. Then we get the optimal choice for $\Delta$ is of the form $\Delta = c \delta^{1/2}\times(\ell(\theta, \xi) - \EE_{\PP}[\ell(\theta, \xi)])$ due to centering implied by $\EE_{\PP}[\Delta] = 0$, where the normalizing constant $c$ can be directly computed from $\EE_{\PP}[|\Delta|^2] = 2 \delta / \phi''(1)$. Plugging in this choice of $\Delta$, we obtain the aforementioned result. 

The most important insight in this analysis is the form of the reweight employed by the adversary. In particular, the adversary increases the importance of samples for which the loss is large compared to the mean loss and decreases the importance of samples for which the loss is low compared to the mean loss. 

Further, as we can see from the previous result,
$\phi$-divergence DRO 
implicitly considers the bias-variance trade-off, where we identify the bias with the empirical loss $\EE_{\hat{\PP}_n}[\ell(\theta, \xi)] $ and the variance with the empirical variance $\text{Var}_{\hat \PP_n}(\ell(\theta, \xi))$. While directly minimizing the bias with variance regularization may result in a non-convex optimization problem, the DRO formulation keeps the convexity and thus enjoys the tractability~\citep{duchi2018variance}. 

The view of adversarial reweighting sheds light on the adaptive reweighting strategy when we use the gradient descent algorithm to solve the DRO problem~\eqref{eq:DRO}. At each step $t$, the gradient with respect to the parameter $\theta$ at $\theta_t$ is computed as:
\begin{align*}
    \frac{\partial}{\partial \theta}\Bigg|_{\theta =\theta_t} ~\sup_{\QQ \in {\mc B}_{\delta}(\hat \PP_n)} \EE_{\QQ}[\ell(\theta, \xi)] = \sum_{i=1}^{n} \omega_{i}^{\star} \frac{\partial \ell}{\partial \theta}(\theta_t, \xi_i),
\end{align*}
where 
\begin{align*}
    & \boldsymbol{\omega}^\star = (\omega^{\star}_1, \dots, \omega^{\star}_n)^\top = \mathop{\arg\max}_{\boldsymbol{\omega} \in C_\delta} \sum_{i=1}^{n} \omega_i \ell(\theta_t, \xi_i),\\
    & C_\delta = \left\{\boldsymbol{\omega} = (\omega_1, \dots, \omega_n)^{\top}:~\frac{1}{n}\sum_{i=1}^{n}\phi(n \omega_i) \leq \delta,~
    \sum_{i=1}^{n}\omega_i = 1, ~\omega_i \geq 0,~\forall i \in [n]\right\}.
\end{align*}
Here, we assume the optimal solution of $\omega^\star$ is unique and $\ell$ is differentiable in $\theta$ for simplicity, and in general, the subgradient  can be computed instead of the aforementioned gradient. This type of adaptive reweighting is well-known in popular machine learning algorithms, like AdaBoost~\citep{freund1997decision}. Actually, the AdaBoost algorithm can be recovered in the DRO framework~\citep{blanchet2019distributionally}.

\subsubsection{Optimal Transport-Based DRO}Another natural approach in modeling discrepancies in probability distributions is based on perturbing the actual random outcomes (as opposed to the likelihood of the outcome, as in the $\phi$-divergence case). This approach can be made operational using the optimal transport discrepancies, the corresponding duality theory has been developed in, for example, studies such as \citet{mohajerin2018data,blanchet2019robust, gao2023distributionally}. This line of research has provided a probabilistic interpretation of various forms of regularization, more precisely, so-called norm regularization which is a widely adopted strategy to effectively address the issue of overfitting in machine learning models \citep{tibshirani1996regression, ng2004feature}. We now provide the definition of an optimal transport discrepancy.  

\begin{definition}[Optimal transport discrepancy]\label{def:ot} Assume that $c: \Xi \times \Xi \rightarrow [0, +\infty]$ is a nonnegative lower semi-continuous function, 
the optimal transport discrepancy between $\QQ$ and $\PP$ is defined as
\begin{align*}
    & \mathds{D}_c(\QQ, \PP) = \min_{\pi}\left\{\int_{\Xi \times \Xi} c(\xi, \xi') \diff \pi(\xi, \xi'):~ \pi_{\xi} = \QQ, ~\pi_{\xi'} = \PP  \right\},
\end{align*}
where $\pi\in\mc P(\Xi \times \Xi)$ is a coupling of $\QQ$ and $\PP$, and $\pi_{\xi}$ (resp. $\pi_{\xi'}$) denotes the marginal distribution of $\pi$ on $\xi$ (resp. $\xi'$).
\end{definition}

The optimal transport discrepancy has a rich tradition in a wide range of areas in engineering and applied mathematics; see, for example, \citet{villani2009optimal, santambrogio2015optimal,peyre2019computational}. When the cost function is a metric, the optimal transport discrepancy recovers the Wasserstein distance. The choice of the cost function can be used to recover both the weak convergence topology (e.g., $c(\xi,\xi') = \|\xi-\xi'||_2/(1+\|\xi-\xi'\|_2)$  when $\Xi$ is a subspace of the Euclidean space) and the total variation topology (e.g., $c(\xi,\xi') = \mathbb{I}_{\xi \neq \xi'}$). So, statistically speaking, optimal transport provides a flexible approach to compare statistical distributions.

\begin{figure}[ht]
    \centering
    \includegraphics[width=0.6\textwidth]{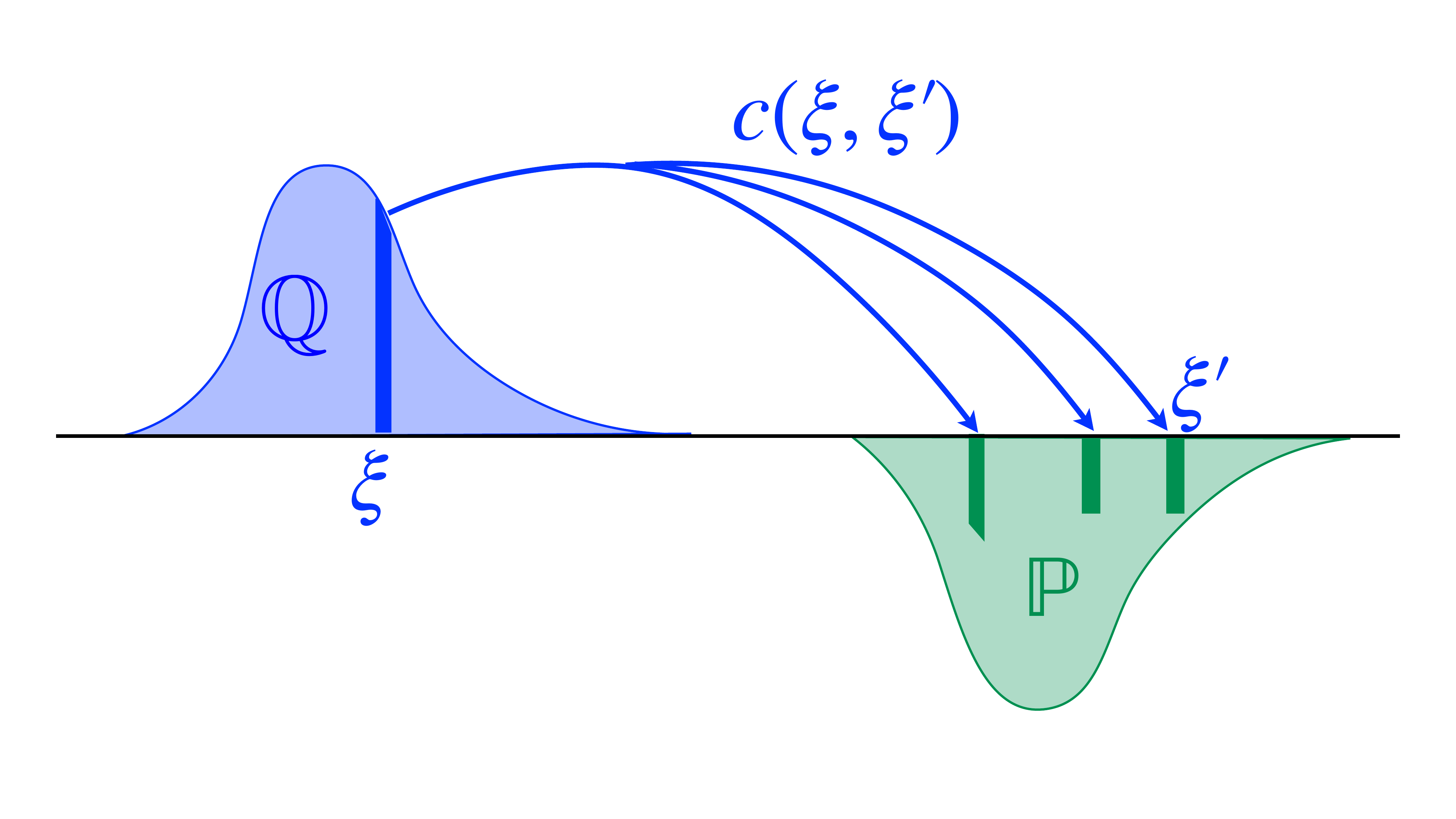}
    \vspace{-2mm}
    \caption{Illustration of optimal transport. The source of mass is denoted by $\QQ$ and the target is denoted by $\PP$. The optimal transport discrepancy between $\QQ$ and $\PP$ is the minimal total cost incurred by transporting all of the mass from $\QQ$ (source) to $\PP$ (target).}
    \label{fig:ot}
\end{figure}

The intuition in the definition of the optimal transport discrepancy is that there is a source of mass, say, a pile of sand, and a target, say, a sinkhole, which are described by two distributions $\PP$ and $\QQ$, respectively. The cost per unit of mass transferred from location $\xi$ in the pile of sand to location $\xi'$ at the sinkhole is given by $c(\xi,\xi')$. The objective function reflects the total cost incurred by transporting all of the mass and the constraints reflect that the profile of the pile of sand is modeled by $\PP$ and the profile of the sinkhole is modeled by $\QQ$, this is illustrated in the Figure~\ref{fig:ot}.

By employing optimal transport as a way to explore the impact of distributional uncertainty, we are intuitively modeling an environment in which the adversary is allowed to act as a transporter of mass, moving the points around subject to budget constraint in the expected cost of transporting mass to achieve maximum impact in the expected loss for the resulting mass configuration.

As we now illustrate, optimal transport discrepancy-based DRO is known to relate to norm regularization in many scenarios. For example, the next result shows that this approach can {\it{exactly}} recover the square-root LASSO estimator introduced in~\citet{belloni2011square}.
\begin{theorem}[Squre-root LASSO {\citep[Theorem 1]{blanchet2019robust}}]\label{thm:lasso}
We assume the random input takes the form $\xi_i = (x_i, y_i) \in \R^{d+1}$, the loss function is $\ell(\theta, \xi) = (y - \theta^{\top} x)^2$, and the cost function admits
\begin{align*}
    c((x,y), (x', y')) =
     \left\{
    \begin{array}{lc}
       \|x - x'\|_q^2  &\quad \text{if}~~ y = y', \\
        +\infty &\quad\text{otherwise}. 
    \end{array}
    \right.
\end{align*}
Then, we have
\begin{align*}
    \min_{\theta \in \Theta} \sup_{\QQ \in {\mc B}_{\delta}(\hat \PP_n)} \EE_{\QQ}[\ell(\theta, \xi)] 
    = \min_{\theta  \in \Theta} \left( 
\sqrt{ \EE_{\hat{\PP}_n}[\ell(\theta, \xi)]} + \sqrt{\delta} \|\theta\|_p \right)^2,
\end{align*}
where ${\mc B}_{\delta}(\hat \PP_n) = \{\QQ: \mathds{D}_c(\QQ, \hat \PP_n) \leq \delta\}$ and 
$\frac{1}{p} + \frac{1}{q} = 1$.
\end{theorem}

The optimal transport discrepancy-based DRO extend beyond regression and encompass norm regularization for generalized linear classification models, such as logistic regression and support vector machines with hinge loss.  


\begin{theorem}[$\ell_p$-norm regularization {\citep[Theorem 4, Theorem 14]{shafieezadeh2019regularization}}]
    We assume the random input takes the form $\xi_i = (x_i, y_i) \in \R^{d+1}$. The loss function is set to be: 1. $\ell(\theta, \xi) = L(y \cdot \theta^\top x)$, or 2. $\ell(\theta, \xi) = L(y - \theta^\top x)$, for a continuous function $L$ with Lipschitz constant $1$, i.e., $\sup_{x \neq y} \frac{|f(x) - f(y)|}{\|x - y\|_2} = 1$, and 
\begin{align*}
    c((x,y), (x', y')) =
     \left\{
    \begin{array}{lc}
       \|x - x'\|_q  &\quad y = y' \\
        +\infty &\quad y \neq y'. 
    \end{array}
    \right.
\end{align*}
Then, we have
\begin{align*}
    \min_{\theta \in \Theta} \sup_{\QQ \in {\mc B}_{\delta}(\hat \PP_n)} \EE_{\QQ}[\ell(\theta, \xi)] 
    = \min_{\theta  \in \Theta} \EE_{\hat{\PP}_n}[\ell(\theta, \xi)] + \delta\|\theta\|_p,
\end{align*}
where ${\mc B}_{\delta}(\hat \PP_n) = \{\QQ: \mathds{D}_c(\QQ, \hat \PP_n) \leq \delta\}$ and $\frac{1}{p} + \frac{1}{q} = 1$.
\end{theorem}

Similar norm regularization on the model parameter $\theta$ can also be recovered in the group LASSO~\citep{blanchet2017distributionally}. Beyond recovering the well-established techniques in statistics and machine learning, the optimal transport-based DRO also provides a principled approach to potentially enhance the existing learning methods (see, e.g., \citet{sinha2018certifying}). For a general loss function, the optimal transport-based DRO asymptotically recovers the so-called variation regularization~\citep{gao2022wasserstein}. 

\begin{figure}[ht]
    \centering
    \includegraphics[width=0.9\textwidth]{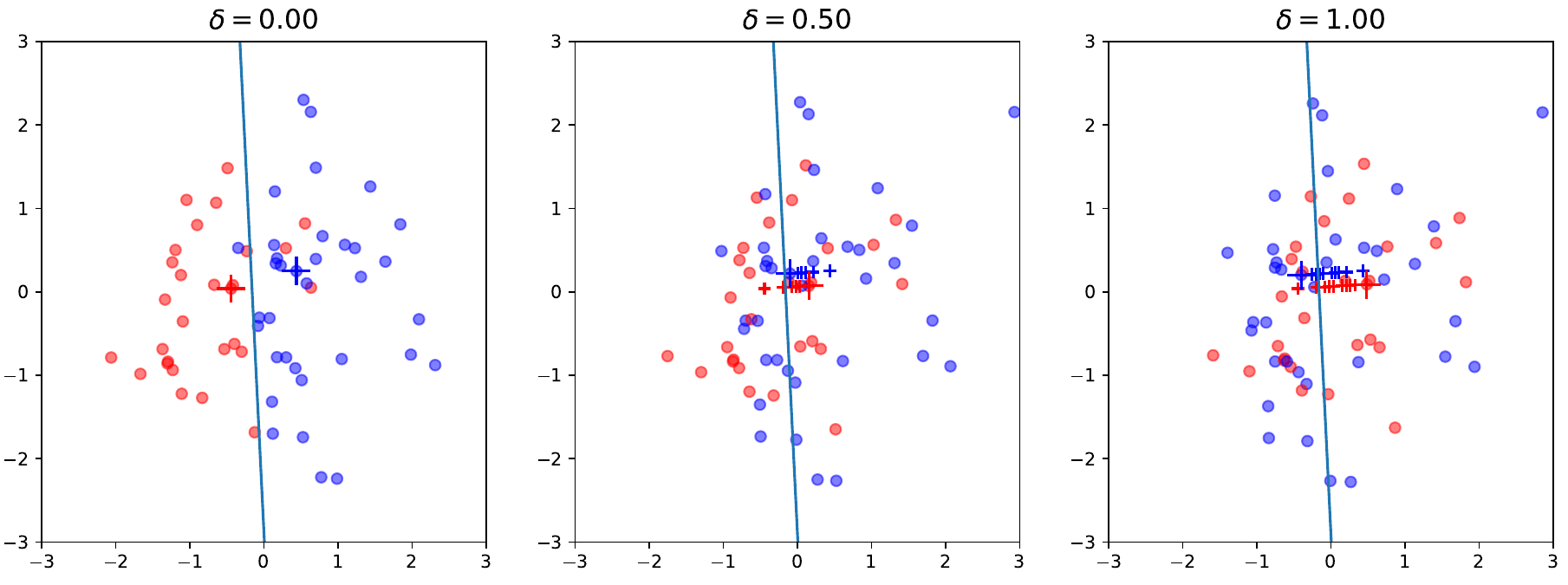}
    \caption{Decision boundary and worst-case distribution. One point from each class is selected, and we use a big + to mark its position. We also use a small + to mark its previous positions when $\delta$ is smaller to visualize its trajectory.}
    \label{fig:worst-case}
\end{figure}

The asymptotic connection to variation regularization can be developed using an intuitive approach similar to the reasoning employed in the $\phi$-divergence setting. It is useful to carry out this analysis because it uncovers the structure of the worst-case adversarial strategy and the appearance of the dual norms as regularization terms. We follow the strategy in the perturbation analysis in~\citet{bartl2021sensitivity}, which provides a rather complete study not only of the optimal adversarial perturbation but also of the optimal parameter selection (see also the analysis in \citet{blanchet2022optimal}). Consider a general formulation in which the adversary maximizes $\EE_{\QQ}[\ell(\theta,\xi)]$ over $\QQ$ such that $\mathds{D}_c(\QQ, \PP) \leq \delta$. Assume, for example, that $c(\xi,\xi')=\|\xi-\xi'\|_q^2$ when $\Xi$ is a subspace of the Euclidean space. The problem for the adversary is equivalent to $\max_{\Delta} \EE_{\PP}[\ell(\theta,\xi + \Delta)]$ over random variables $\Delta$ satisfying $\EE_{\PP}[\|\Delta\|_q^2] \leq \delta$. By applying a Taylor expansion on the objective function, we see that when $\delta$ is small, the adversary is effectively maximizing $\EE_{\PP}[\nabla \ell(\theta,\xi)\Delta]$, where $\nabla$ is the gradient with respect to $\xi$ (in linear models, the parameter will naturally appear by the chain rule thus leading to norm regularization). It is then direct from this observation that $\Delta$ will be chosen by the adversary ``parallel'' or ``aligned'' to $\nabla \ell(\theta,\xi)$ in the corresponding geometry induced by the cost function with the intent of maximizing the loss. This alignment condition is given the dual pair which achieves equality in Holder's inequality. This explains why when the cost is chosen based on the $l_q$ norm, the regularization involves the dual $l_p$ norm. Thus, the perturbation direction employed by the worst-case adversary is fully dictated by the norm dual vector of the loss's gradient with respect to the source of randomness. The size of the norm is fully dictated by the budget constraint $\EE_{\PP}[\|\Delta\|_q^2] \leq \delta$. For example, if the cost  function $c(\xi,\xi')$ is locally quadratic around the diagonal, then $\|\Delta\|_{q} = O(\delta^{1/2})$. As an illustrative example, \citet[Section 4.1.2]{blanchet2022optimal} computes the worst-case adversarial distribution in the setting of logistic regression. Figure~\ref{fig:worst-case} shows the decision boundaries of binary classification based on the DRO solutions with increasing uncertainty budget $\delta$, and the worst-case adversarial distribution after perturbing the empirical distribution. The perturbation trajectory of one sample point from each class is marked by $+$'s. We note that some sample points that are close to the decision boundary (see the markers $+$) will flip their sign of label after adversarial perturbation. As a concrete application, \citet{shafieezadeh2023new} proposes to generate adversarial examples from the worst-case adversarial distribution.

It is useful to contrast the variance regularization of $\phi$-divergence-based DRO compared to the variation regularization which is defined as the norm of the gradient of the loss function with respect to the source of randomness. 

Not surprisingly, given the above analysis, the idea of transporting the sample to its neighborhood has direct connections to adversarial training methods~\citep{goodfellow2014explaining, madry2017towards}. This connection has been directly exposed in \citet{sinha2018certifying}. Interested readers are referred to the tutorials \citet{kuhn2019wasserstein, blanchet2021statistical} for extensive discussion on the optimal transport-based approach.

Some other versions of the optimal transport-based DRO are explored. For example, the martingale constraints are added to the optimal transport-based uncertainty set~\citep{li2022tikhonov, lotidis2023wasserstein} - this means imposing the constraint that the adversary and the baseline models form a martingale pair coupling. By Strassen's theorem~\citep{strassen1965existence}, it implies that the worst-case adversary dominates the baseline model in convex ordering, thus is a constraint that is sensible when the loss is convex as a function of the random noise. Adding sensible constraints to DRO formulations, in turn, is useful to control over-conservative estimators. In the particular case of martingale constraints, it is shown in~\citet{li2022tikhonov} that these constraints recover the classical ridge regression and Tikhonov regularization. A popular variant of the optimal transport discrepancy, the Sinkhorn divergence, is also extensively used in machine learning tasks. This notion can also be used to construct distributional uncertainty sets as demonstrated in \citet{wang2021sinkhorn, dapogny2023entropy, azizian2023regularization}. 

As we shall discuss next, together with $\phi$-divergence and optimal transport can be unified under the lens of optimal transport by adding moment constraints related to martingale constraints. A recent unification of these DRO frameworks is presented in \citet{blanchet2023unifying}, which proposes to lift the sample space by considering the ``likelihood'' itself as a random variable and therefore is amenable to perturbations based on optimal transport, subject to the constraint that a likelihood ratio must have expectation equal to unity. The constraint that likelihood is non-negative can also be handled using optimal transport because the value of ground transportation cost function can be taken as positive infinity in part of its domain (note that only lower semi-continuity is required in the definition). More precisely, the authors consider the lifted empirical measure in the space of $\Xi \times \R_{+}$: $\hat \PP_n \times \delta_1 = ((\xi_1, 1), \dots, (\xi_n, 1))$, the uncertainty set is formulated as ${\mc B}_{\delta}(\hat \PP_n) = \{\QQ \in \mc P(\Xi \times \R_{+}): \mathds{D}_M(\QQ, \hat \PP_n) \leq \delta\}$, where, for a function $\phi(\cdot)$ defined in Definition~\ref{def:phidiv} and a cost function $c(\cdot, \cdot)$ defined in Definition~\ref{def:ot},
\begin{align*}
    & \mathds{D}_M(\QQ, \hat \PP_n) = 
    \min_{\pi}\left\{\int_{(\Xi \times \R_{+}) \times (\Xi \times \R_{+})}\bar c((\xi, u), (\xi', u')) \diff \pi((\xi, u), (\xi', u')): \right. \\
    & \qquad\qquad\qquad\qquad\left. \pi_{(\xi, u)} = \QQ, ~\pi_{(\xi', u')} = \hat \PP_n\times  \delta_1, \EE_{\pi}[u] = 1\right\},\\ 
    & \bar c((\xi, u), (\xi', u'))  = u \cdot c(\xi, \xi') + (\phi(u) - \phi(u'))^{+}.
\end{align*}
This particular choice of a cost function interpolates the aforementioned optimal transport-based and $\phi$-divergence-based methods. Moreover, the resulting unified DRO problem is formulated as
\begin{align*}
    \min_{\theta \in \Theta} \sup_{\QQ \in {\mc B}_{\delta}(\hat \PP_n)} \EE_{\QQ}[u \cdot \ell(\theta, \xi)],
\end{align*}
where the modified expected loss $\EE_{\QQ}[u \cdot \ell(\theta, \xi)]$ illustrates the lifting technique that introduces the variable $u$ as a ``likelihood''.

This approach provides a principled way to interpolate the norm regularization and the variance regularization. Moreover, the worst-case adversarial strategies simultaneously re-weight and perturb actual likelihoods based on dual gradient directions as discussed earlier in the settings of pure $\phi$-divergence-based and optimal transport-based DRO, respectively. Moreover, if one alternatively chooses the nominal distribution with a kernel density, then this framework can also recover the Sinkhorn divergence-based DRO.


\subsubsection{Integral Probability Metric-Based DRO}
In statistics, energy distances and maximum mean discrepancies (MMD) are also widely considered~\citep{szekely1989potential}. They also arise in the framework of kernel-based DRO~\citep{staib2019distributionally, zhu2021kernel}. 
\begin{definition}[Maximum mean discrepancy]
For a reproducing kernel Hilbert space $\mc H$ (see, e.g.~\citet{berlinet2011reproducing}), the maximum mean discrepancy between distributions $\QQ$ and $\hat \PP_n$ is defined as
\begin{align*}
    \mathds{D}_{\mc H}(\QQ, \hat \PP_n) = \sup_{f:\|f\|_{\mc H} \leq 1}~\int_{\Xi} f \diff \QQ(\xi) - \int_{\Xi} f \diff \hat \PP_n(\xi).    
\end{align*}
\end{definition}
When choosing different spaces $\mc H$, the kernel-based DRO can recover various DRO formulations, for example, moment-constraint DRO (see \citet[Example 3.4]{zhu2021kernel}). Thus, this type of formulation provides a unified approach to studying a class of uncertainty sets in the DRO. A more general DRO framework is to consider the integral probability metrics (IPM) as the statistical distance (see, e.g., \citet[Corollary 3.1.1]{zhu2021kernel}).
\begin{definition}[Integral probability metrics]
    For a function class $\mc F$, the integral probability metric between distributions $\QQ$ and $\hat \PP_n$ is defined as
    \begin{align*}
        \mathds{D}_{\mc F}(\QQ, \hat \PP_n) = \sup_{f\in\mc F}~\int_{\Xi} f \diff \QQ(\xi) - \int_{\Xi} f \diff \hat \PP_n(\xi).
    \end{align*}
\end{definition}
Specifically, assume that $\Xi = \R^{d}$, then
\begin{itemize}
    \item if $\mc F = \{f: \|f\|_{\mc H} \leq 1\}$ for a Hilbert space $\mc H$, $\mathds{D}_{\mc F} = \mathds{D}_{\mc H}$;
    \item if $\mc F = \{f: \|f\|_{Lip} \leq 1\}$, where
    \begin{align*}
        \|f\|_{Lip} = \sup_{\xi,\xi'} \frac{|f(\xi) - f(\xi')|}{\|\xi - \xi'\|_q},
    \end{align*}
    then $\mathds{D}_{\mc F} = \mathds{D}_{c}$ for $c(\xi,\xi') = \|\xi - \xi'\|_q, q\geq 1$, which results from the Kantorovich-Rubenstein duality~\citep{kantorovich1958space};
    \item if $\mc F = \{f: \|f\|_{\infty} \leq \frac{1}{2}\}$, then $\mathds{D}_{\mc F} = \mathds{D}_{\phi}$ for $\phi(z) = \frac{1}{2}|z - 1|$, which results from the variational representation of the $\phi$-divergence. 
\end{itemize}

\subsubsection{Calibration of DRO with Side Information}

In practical scenarios, available side information often includes: (i) the structure of the data-generating distribution; (ii) the characteristics of the out-of-sample environment; and (iii) domain knowledge. This information is especially helpful in mitigating the conservatism of the DRO while robustifying the DRO solution against the unknown out-of-sample environment. In the subsequent discussion, we showcase several DRO formulations that leverage diverse types of information, illustrating potential integrations of DRO with specific side information—all grounded in the aforementioned DRO formulations.

(i) \textit{Correlation inside data.} When the observed data have internal correlation within its structure, the DRO framework can be applied with corresponding structure of uncertainty sets. For example, given i.i.d.~samples $(x_i, y_i), i=1,\dots,n$ generated from $\PP_\star$, where $y_i \in \R^m$ represents the response variable and $x_i\in \R^n$ represents the predictor or covariate. Assume that the practitioner wants to solve
\begin{align*}
    \min_{\theta \in \Theta} \EE_{\PP_\star}\left[\ell(\theta, y) | x = x_0\right]\quad \text{for $x_0 \in \R^n$,}
\end{align*}
which provides an estimator of $y$ exploiting the side information carried by $x$. This problem presents several challenges. Firstly, the probability distribution $\PP_\star$ is typically unknown, and only i.i.d. samples are available, which may be at risk of contamination. Additionally, there may be a scarcity of observations in the samples with covariate value $x = x_0$. To address these challenges without relying on additional parametric assumptions, \cite{nguyen2020distributionally, nguyen2021robustifying} consider the DRO formulation 
\begin{align*}
    \min_{\theta \in \Theta} \sup_{\substack{\bar \PP \in {\mc B}_{\delta}(\hat \PP_n), \\ \bar \PP(x \in \mc N(x_0)) > 0}} \EE_{\bar \PP}\left[\ell(\theta, y) | x \in \mc N(x_0)\right],
\end{align*}
where ${\mc B}_{\delta}(\hat \PP_n) = \left\{ \PP: \mathds{D}_c(\PP, \hat \PP_n) \leq \delta \right\}$ and $\hat \PP_n = \frac{1}{n}\sum_{i=1}^{n}\delta_{(x_i, y_i)}$. Here, the side information is specifically modelled
as a neighborhood $\mc N(x_0)$ around a covariate $x_0$ and the distribution $\bar \PP$ is constraint such that $\bar \PP(x \in \mc N(x_0)) > 0$ to avoid conditioning on a set of measure zero. When $\delta = 0$ and $\mc N(x_0)$ exactly contains the $k$ nearest samples in $(X_i, 1\leq i\leq n)$ from $x_0$, then this formulation recovers the $k$-nearest neighbor regression estimator, which is consistent~\citep{stone1977consistent} when $k$ is suitably chosen with respect to $n$. Thus, this formulation can also be considered as a robustification of the $k$-nearest neighbor estimator.

As another example, \citet{hu2018does} and \citet{sagawa2019distributionally} leverage the prior knowledge of correlations between samples in the training data to robustify their decision against potential group shifts, formally assuming that the data-generating distribution is a mixture of $m$ latent groups, i.e.,
\begin{align*}
    \PP_{\star} = \sum_{\eta = 1}^{m} q_{\eta}\PP_{\xi | \eta}.
\end{align*}
Correspondingly, the uncertainty set of the group DRO is built on the latent probability $q_{\eta}$ using $\phi$-divergence, i.e.,
\begin{align*}
    \mc B_{\delta}(\hat \PP_n) = \{\QQ: \mathds{D}_{\phi}(\QQ, \hat \PP_n) \leq \delta, \QQ_{\xi | \eta} = (\hat \PP_n)_{\xi | \eta}\ \forall \eta = 1, \dots, m \},
\end{align*}
where the distributions $\QQ$ and $\hat \PP_n$ are assumed to have the same structure with $m$ latent groups.

Additional formulations motivated by group regularization in the context of optimal transport-based DRO are studied in \citet{blanchet2017distributionally}.

(ii) \textit{Out-of-sample environmental shift.} The uncertainty set characterizes the potential shifts in the out-of-sample environment or, put differently, envisions how an adversary might perturb the environment upon deployment of practitioners' decisions. Contrasting with examples in the preceding paragraphs, where adversarial perturbations are assumed to be linked to the internal structure of the data-generating distribution, practitioners can also directly leverage their knowledge of potential distributional shifts to construct uncertainty sets.

An instance of reshaping or informing the uncertainty set with prior knowledge of potential distributional shifts is illustrated in~\citet{blanchet2022optimal}. In the portfolio selection problem, the return vector $\xi$ is the random input of the optimization problem. The authors construct an uncertainty set on the distribution of $\xi$ using an optimal transport cost (Definition~\ref{def:ot}) which is informed by current market information, in particular, the implied volatility derived by calibrating option prices based on the Black-Scholes formula with the market option prices. The implied volatility provides insight into the market participant's collective future belief about the volatility of an asset. Thus, if the implied volatility is large, the transportation cost should be relatively low so that the adversary can more efficiently use the budget to explore the adversarial impact of investing in an asset with, for instance, high implied volatility. Likewise, if the implied volatility is small, it is sensible to impose a relatively high transportation cost because in this way the DRO formulation will discourage the adversary from exploring the impact of future variations on assets that are perceived to be safe collectively by the market, which is captured by the implied volatility. Therefore, the formulation illustrated in \citet{blanchet2022optimal} takes the following form.

\begin{align*}
    &~{\mc B}_{\delta} = \{\QQ: \mathds{D}_{c}(\QQ, \hat \PP_n) \leq \delta\}, \\
    &~c(\xi_i, \xi) = (\xi_i - \xi)^{\top} A_i (\xi_i - \xi),\\
    &~A_i = \frac{\bar V}{V_i} I_{d}\quad i = 1,\dots, n.
\end{align*}
Here, the transport cost of perturbing the sample return $\xi_i$ is defined using the squared Mahalanobis distance with a specified matrix $A_i$. $A_i$ is the identity matrix scaled by $\frac{\bar V}{V_i}$, where $\bar V = \frac{1}{n}\sum_{i=1}^{n}V_i$ and $V_i$ is the implied volatility corresponding to the sample return $\xi_i$. As a result, for the distributions in the uncertainty set $\mc B_{\delta}$, it is cheaper to perturb sample returns with higher implied volatility. The formulation illustrates the intuition discussed earlier, namely, that higher implied volatility suggests larger price uncertainty in future returns by the collective market. 

Similarly in~\citet{blanchet2021doubly}, the authors tune the cost function defined in the optimal transport discrepancy to apply in classification problems. The authors first fit a cost function with the property that observations with the same labels are close, while observations with different labels are far apart. Then, this cost function is used as a transportation cost in an optimal transport-based DRO logistic regression formulation. The intuition is that the adversarial budget will be invested more efficiently if the adversary is encouraged to perturb data points that are easier to be flipped in the decision boundary, i.e., moved to the population with the opposite label. 

Yet another example arises in the context of unsupervised learning \citep{blanchet2020semi, blanchet2021sample}, where the authors unlabeled observations to shape the distributional uncertainty set. The intuition is that the underlying data may lie in a lower dimensional space and only variations along such a lower dimensional space should be explored by the adversary in an optimal transport formulation.

(iii) \textit{Domain knowledge.} The utilization of domain knowledge to enhance the out-of-sample performance of decisions has been extensively investigated in the field of domain adaptation (refer to, e.g., \citet{weiss2016survey, wilson2020survey, iman2023review}). As a robust formulation aimed at minimizing worst-case risk under potential distributional shifts, the DRO framework can be synergistically integrated with the concept of domain adaptation to further enhance its out-of-sample performance. 

For instance, \citet{taskesen2021sequential} proposes two strategies for applying the DRO in the context of domain adaptation to solve the linear prediction problem. The authors introduce the DRO in a parametric setting such that the uncertainty set is prescribed using only finite parametrization of distributional moments. Here, for simplicity of notation, we present the high-level idea of their work without explicitly delving into the details of their parametric setting. Given nominal distribution from source and target, namely, $\hat \PP_{S}$ and $ \hat\PP_{T}$, the authors propose to create a class of ``experts'' that integrates the knowledge of source and target and then aggregate the experts. To create robust experts enhancing its out-of-sample performance, they consider two types of uncertainty sets $\mc B$ and solve the distributionally robust least squares estimation problem in the following form.
\begin{align*}
    & \quad \min_{\theta \in \Theta} \sup_{\QQ \in \mc B} \EE_{\QQ}\left[(\beta^{\top} x - y)^2\right],~ \text{where} \\ 
    &  
    \left.
    \begin{array}{l}
        (a)~ \mc B =\mc B_{\delta}(\hat \PP_{\lambda}) = \{\QQ: \mathds{D}(\QQ, \hat \PP_{\lambda}) \leq \delta\};  \\
        (b)~ \mc B = \mc B_{\delta_S}(\hat \PP_{S}) \bigcap \mc B_{\delta_T}(\hat \PP_{T}) = \{\QQ: \mathds{D}(\QQ, \hat \PP_{S}) \leq \delta_{S}, \mathds{D}(\QQ, \hat \PP_{T}) \leq \delta_{T}\}.
    \end{array}
    \right.
\end{align*}
Here, the radii $\delta, \delta_S, \delta_T > 0$, $\hat \PP_{\lambda}$ is an interpolation of $\hat \PP_{S}$ and $\hat\PP_{T}$ parametrized by $\lambda \in [0,1]$, and the statistical divergence $\mathds{D}$ can be taken as $\phi$-divergence $\mathds{D}_{\phi}$ (Definition~\ref{def:phidiv}) or optimal transport discrepancy $\mathds{D}_{c}$ (Definition~\ref{def:ot}). Thus, for each type (i.e. $(a)$ and $(b)$) of uncertainty and each statistical divergence $\mathds{D}$, the authors choose various values of radii and parameters to generate a class of ``experts'' that integrates the knowledge from source and target.

Another example for solving linear regression in the context of domain adaptation is proposed in \citet{zhang2022class}. In their work, the authors evaluate the worst-case risk of the estimator over all pairs of source and target within a specified distance between each other. This formulation can also be viewed as a DRO problem such that the source lies in the uncertainty set centered at the target, which is prescribed as a neighborhood ball in the distribution space. Notably, \citet{zhang2022class} also discusses the minimax optimality of the proposed estimator with respect to all pairs of source and target within the specified distance in a non-asymptotic sense.



\subsection{Statistical Properties of DRO}\label{sec:statguarantee}

In this subsection, we focus on the statistical properties of the estimator obtained by solving the DRO problem, referred to as the DRO estimator in the following. Our focus is on the large sample regime, where the sample size approaches infinity while the dimension of individual samples remains fixed. We will discuss the limiting behavior of the DRO estimator to characterize its efficiency in addition to its finite sample properties. 

In the scenario when the practitioner wants to defend against potential overfitting, it is natural to shrink the radius of the uncertainty set in the DRO when the number of sample goes to infinity, i.e., $\lim_{n\rightarrow \infty}\delta_n = 0$. This is because the empirical distribution $\hat \PP_n$ converges to the data-generating distribution $\PP_{\star}$ and thus the uncertainty with respect to $\PP_{\star}$ decays to zero. Following this vein, a series of studies explores the appropriate shrinkage rate of the radius of the uncertainty set such that the DRO estimator is consistent and nearly efficient. These studies also give rise to new tools for statistical inference, such as confidence region construction.

Alternatively, if the practitioner aims to provide a robust estimator that performs well not only under the data-generating distribution but also in the face of the unknown distributional shift—a common modeling challenge—the determination of the radius becomes a nuanced task. In this scenario, the radius may not solely hinge on the sample size but also on the decision maker's risk posture concerning potential variations in the distribution, particularly relative to the empirical distribution derived from past observations. From this perspective, it is non-trivial (and not purely a statistical problem) to optimally choose not only the size of the distributional uncertainty but the distributional uncertainty set. Some studies thus focus on fixing an uncertainty set and consider and compare the DRO solution using empirical data with its population version, which is the solution of the same DRO problem except for centering its uncertainty set at the true data-generating distribution.  

In the subsequent discussion, we focus on the uncertainty sets in the form previously described:
\begin{align*}
    {\mc B}_{\delta_n}(\hat \PP_n) = \left\{\QQ: \mathds{D}(\QQ, \hat \PP_n) \leq\delta_n\right\},
\end{align*}
where $\mathds{D}$ represents a probability metric to be chosen, and the radius $\delta_n$ may depend on $n$. The statistical properties of other DRO variants featuring more general uncertainty sets have, to a large extent, remained underexplored.

\subsubsection{{Selection of Radius $\delta_n$}}\label{sec:selectradii}
The selection of the radius $\delta_n$ significantly impacts the statistical property of the DRO solution. When $\delta_n$ is too large, the resulting strategy is too conservative to learn useful information from the data. When too small, the strategy is not robust enough to combat against the noise in the data. In many scenarios, the radius $\delta_n$ can be directly interpreted as the penalization parameter (see, for example, Theorem~\ref{thm:var-reg}, \ref{thm:lasso}). This connection—the penalization parameter is the radius of uncertainty set—provides an interpretable way of selecting the penalization parameter from the view of uncertainty magnitude, other than possibly computation-intensive machine learning methods like cross-validation. 

From a theoretical perspective, a sequence of studies suggests ensuring that $\PP_\star \in {\mc B}_{\delta_n}(\hat \PP_n)$ with high probability (for example, \citet[Theorem 3.5]{mohajerin2018data}). However, for the likelihood-based $\phi$-divergence, $\PP_\star$ may never be in the uncertainty set ${\mc B}_{\delta_n}(\hat \PP_n)$ when it has continuous density because in this case the likelihood ratio between $\PP_\star$ and the empirical distribution $\hat \PP_n$ is not well-defined. To address this issue, one approach is to consider the parametric distributions that are compatible with $\hat \PP_n$ and ensure that the uncertainty set contains at least one of them with high probability (see, e.g.~\citet[Corollary 4]{delage2010distributionally}, \citet[Lemma 4.4]{nguyen2020distributionallypara}).

For another popular statistical distance, optimal transport discrepancy, the choice of radius may be subject to the curse of dimensionality. In particular, the optimal transport discrepancy between $\PP_\star$ and $\hat \PP_n$ with cost function $c(\xi, \eta) = \|\xi - \eta\|^p, \xi, \eta \in \R^d$ is in the order of $n^{-p/d}$ in expectation \citep[Theorem 1]{fournier2015rate}, which implies that the radius $\delta_n$, if selected in the aforementioned way, will also grow in the order of $n^{-p/d}$ and result in a too conservative learning strategy in practice. To overcome this issue, instead of letting the uncertainty set contain $\PP_\star$ with high probability or fitting $\hat \PP_n$ to parametric distributions, another radius selection method is to guarantee that the uncertainty set contains at least one distribution of which the learned parameter is compatible with $\PP_\star$. In particular, let
\begin{align*}
    &~\theta(\QQ) = \mathop{\arg\min}_{\theta\in\Theta} \EE_{\QQ}[\ell(\theta, \xi)], \\
    &~\Lambda_{\delta_n} = \left\{\theta(\QQ): \QQ \in {\mc B}_{\delta_n}(\hat \PP_n)\right\}.
\end{align*}
Then the radius $\delta_n$ is selected such that $\theta(\PP_\star) \cap \Lambda_\delta(\hat \PP_n) \neq \emptyset$ in high probability. Formally, for a specified probability threshold $\alpha$, the radius $\delta_n$ is selected as the smallest number such that
\begin{align}\label{eq:rwpi}
    \PP_{\star}\left(\theta(\PP_\star) \in \Lambda_{\delta_n}(\hat \PP_n) \right) \geq 1- \alpha,
\end{align}
where the first $\PP_{\star}$ is short for the infinite product data-generating distribution $\PP_{\star}^{\otimes \infty}$. While this optimization problem appears daunting, in Section \ref{subsect:confidence} we will explain how to reduce it to a projection which (in the setting of $\phi$-divergence) is closely related to empirical likelihood. 
A concrete recipe for implementing this selection method is presented in~\citet[Algorithm 1]{blanchet2021statistical}. This selection method can be extensively applied to the optimal transport discrepancy-based DRO~\citep{blanchet2019robust, blanchet2019distributionally} and the $\phi$-divergence-based DRO \citep{lam2017empirical,blanchet2019distributionally}. 

Notably, the selection rule \eqref{eq:rwpi} finds its connection in the classic LASSO regression literature. In the high dimension regression setting, Theorem~\ref{thm:lasso} shows that the radius $\delta_n$ of uncertainty set is interpreted as the penalization parameter in front of the norm regularization of the regression parameter. In this case, the rule \eqref{eq:rwpi} suggests a similar selection of $\delta_n$ as in the LASSO literature \citep[Theorem 7]{blanchet2019robust} and it scales similarly in the high dimensional setting and it is also independent of the residual standard errors as in the square-root LASSO setting in~\citet{belloni2011square}.

\subsubsection{Asymptotic Normality of DRO Estimators} \label{Subsubsec:asymnormal}  
In this part, we discuss the limiting behavior the DRO estimator by presenting their asymptotic normality under regular assumptions. We begin by discussing the scenario where $\delta_n$ approaches zero as $n$ tends to infinity, and then we explore the case where $\delta_n$ remains fixed throughout.

Precisely, we define the DRO estimator with shrinking radius $\delta_n$ to be
\begin{align*}
    \hat \theta_n \in \mathop{\arg\min}_{\theta \in \Theta} \sup_{\QQ \in {\mc B}_{\delta_n}(\hat \PP_n)} \EE_{\QQ}[\ell(\theta, \xi)].
\end{align*}
and the parameter to be learned, which is assumed to be unique, is defined as
\begin{align*}
    \theta_{\star} = \mathop{\arg\min}_{\theta \in \Theta} \EE_{\PP_{\star}}[\ell(\theta, \xi)].
\end{align*}

Under regular assumptions, when the radius of the uncertainty set $\delta_n$ shrinks to zero as $n$ grows to infinity, the DRO estimator $\hat \theta_n$ converges to $\theta_{\star}$ almost surely. Intuitively, when $\delta_n$ approaches zero too fast, the limiting behavior of the DRO estimator will be close to the minimizer of the empirical risk without robust approach; when $\delta_n$ grows to slowly, the DRO estimator will be too conservative, leading to poor efficiency when estimating $\theta_{\star}$. It turns out, when appropriately choosing the shrinking rate of $\delta_n$ (as outlined in Section \ref{sec:selectradii}), the error of the DRO estimator with respect to $\theta_{\star}$ can be characterized by the central limit theorem with an expected convergence rate $1/\sqrt{n}$.

Alternatively, the results in Section~\ref{sec:drostattasks} can also help to decide the shrinking rate of $\delta_n$. For example, when the uncertainty set is based on the $\phi$-divergence and $\phi''(1)>0$, Theorem \ref{thm:var-reg} indicates that the error in the estimation of the optimal expected loss is of order $O(\delta_n^{1/2})$. In view of the central limit theorem, this suggests selecting $\delta_n = \frac{\bar \delta}{n}$ for some constant $\bar \delta > 0$. (If $\delta_n$ shrinks faster than $\frac{1}{n}$, the DRO solution cannot provide a satisfying robust certificate as we will discuss in Section \ref{sec:finitesample}). For the constant $\bar \delta$, the procedures outlined in Section \ref{sec:selectradii} and \ref{subsect:confidence} provide its precise selection. As a result, given $\delta_n = \frac{\bar \delta}{n}$, the central limit theorem holds as follows
\begin{theorem}[{\citet{duchi2018variance}}, Theorem 6, informal]
    Under smoothness assumptions on $\ell(\cdot, \cdot)$, 
    \begin{align*}
        \sqrt{n} \left(\hat \theta_n - \theta_{\star}\right) \overset{d}{\rightarrow} N\left(-\sqrt{2\bar \delta} b, \Sigma\right),
    \end{align*}
    for some vector $b$ and matrix $\Sigma$ that depends on $\PP_{\star}, \theta_{\star}, \ell(\cdot,\cdot)$. Here, $\overset{d}{\rightarrow}$ denotes the convergence in distribution.
\end{theorem}

The asymptotic bias is explicitly characterized in \citet{duchi2018variance}, but the point is that both the bias and the covariance matrix depend on $\theta_\star$. This dependence is typically continuous in $\theta_\star$, thus the asymptotically valid confidence regions of $\theta_\star$ can be built based on this result with any consistent plug-in estimator.

A completely analogous result can be obtained in the context of the optimal transport discrepancy. Once again, the choice of $\delta_n$ can either be guided by the method outlined in the section \ref{sec:selectradii} to provide its precise selection, or by using the analysis outlined earlier for the structure of the worst case distribution. For example, when the cost function is $c(\xi,\xi') = \|\xi-\xi'\|_q^2$, we saw that the perturbation size is of order $O(\delta_n^{1/2})$, thus leading to an error in the optimal expected loss of order $O(\delta_n^{1/2})$. This implies once again selecting $\delta_n = \frac{\bar \delta}{n}$, the central limit theorem results in a completely similar format asymptotically normal limit. It is important, however, that the asymptotic bias is different, but both the asymptotic variances (in the contexts of $\phi$-divergence and optimal transport discrepancy) coincide with the case of the empirical risk minimization estimator (i.e. the case $\delta_n=0$). The asymptotic bias is typically also continuous in $\theta_*$ and therefore any consistent plug-in estimator can be used to generate asymptotically valid confidence intervals; see \citet[Theorem 1]{blanchet2022confidence}. The work of \citet{blanchet2023statistical} provides a comprehensive discussion of both the DRO and the $\phi$-divergence asymptotic normality results building from a sensitivity analysis perspective and studying the different asymptotic distributions for different choices of uncertainty radii. 

It is important to note that the DRO estimator has a significant bias of order $O(n^{-1/2})$ compared to the bias of the standard empirical risk minimization estimator. Therefore, it is valid for the reader to wonder what is the point of a DRO-based estimator in the face of the above result which indicates that the asymptotic mean squared error. 

This is a valid criticism from the standpoint of a purely mean-squared error criterion as a function of the sample size, one that we will address in the conclusion section of this review paper. It suffices to say here, that mean-squared error is not the only criterion of interest and that there are additional parameters of interest and not only sample size (e.g. the complexity of the model to be learned) that are important. The optimality and efficiency of DRO estimators as a function of various statistical parameters of interest (modeling complexity class and equitability or fairness, among other criteria) are topics of significant research interest. In the next section, we will review finite sample results which provide insight into some of these questions. 

By contrast, if we keep the radius $\delta_n$ fixed as $n$ goes to infinity, and consider its convergence of the DRO estimator to its population counterpart (defined as follows), we have central limit theorem with zero asymptotic bias. To be precise, we define the DRO estimator with fixed radius $\delta$ to be 
\begin{align}\label{eq:DROfix}
    \hat \theta_{n, \delta} \in \mathop{\arg\min}_{\theta \in \Theta} \sup_{\QQ \in {\mc B}_{\delta}(\hat \PP_n)} \EE_{\QQ}[\ell(\theta, \xi)].
\end{align}
and its population counterpart, which is assumed to be unique, is defined as
\begin{align*}
    \theta_{\star, \delta} = \mathop{\arg\min}_{\theta \in \Theta} 
    \sup_{\QQ \in {\mc B}_{\delta}( \PP_{\star})}\EE_{\QQ}[\ell(\theta, \xi)].
\end{align*}
When the uncertainty set is based on $\phi$-divergence, the central limit theorem of the convergence of $\hat \theta_{n, \delta}$ is given by
\begin{theorem}[\citet{duchi2021learning}, Theorem 11, informal] Under regular assumptions on $\ell$,
\begin{align*}
    \sqrt{n} \left(\hat \theta_{n, \delta} - \theta_{\star, \delta} \right) \overset{d}{\rightarrow} N\left(0, \Sigma^*\right),
\end{align*}
for some variance matrix $\Sigma^*$ depending on $\PP_{\star}, \ell(\cdot,\cdot), \delta, \phi, \theta_{\star, \delta}$.   
\end{theorem}
The underlying idea of proving this result is to view the DRO estimator as an M-estimator in view of the DRO duality problem. To be precise, by \citet[Section 3.2]{shapiro2017distributionally}, the DRO duality problem is
\begin{align*}
    \sup_{\QQ \in {\mc B}_{\delta}(\hat \PP_n)} \EE_{\QQ}[\ell(\theta, \xi)] = \inf_{\lambda \geq 0, \mu \in \R} \left\{\EE_{\hat \PP_n}\left[\lambda \phi^*\left(\frac{\ell(\theta, \xi) - \mu}{\lambda}\right)\right] + \lambda \delta + \mu\right\},
\end{align*}
where $\mc B_{\delta}(\hat \PP_n) = \{\QQ: \mathds{D}_{\phi}(\QQ, \hat \PP_n) \leq \delta\}$, $\phi^*(s) = \sup_{t\in\R}\{s\cdot t - \phi(t)\}$. Similarly, in other context such as the optimal transport discrepancy, the central limit theorem of $\hat \theta_{n, \delta}$ can be established by solving the corresponding DRO duality problem.



\subsubsection{Finite-Sample Guarantees of DRO Estimators}\label{sec:finitesample} In the studies of statistical properties of DRO, many efforts have been made to characterize the out-of-sample generalization power of the DRO estimator, particularly under the true data-generating distribution. These endeavors involve establishing finite-sample statistical guarantees in the form of
\begin{align}\label{eq:certificate}
    \EE_{\PP_{\star}}[\ell(\hat \theta_n, \xi)] \leq \inf_{\theta \in \Theta}\sup_{\QQ \in {\mc B}_{\delta_n}(\hat \PP_{n})} \EE_{\QQ}[\ell(\theta, \xi)]
\end{align}
with high probability (see, for example, \citet[Theorem 3]{delage2010distributionally}, \citet[Theorem 3.5]{mohajerin2018data}). The optimal value of the DRO problem is thus termed as the certificate of the out-of-sample performance of $\hat \theta_n$. The reason for emphasizing this one-sided upper bound is probably that in many applications of the DRO, underestimation of the loss is more harmful than overestimation of the loss.

The more recent studies, such as \citet[Corollary 5]{duchi2018variance}, \citet[Theorem 1]{an2021generalization} and \citet[Theorem 1]{gao2022finite}, address the issue of over-conservatism in DRO by selecting a more refined uncertainty set with smaller radii $\delta_n$ (e.g. by sending $\delta_n$ to zero in a faster rate). This choice often introduces an additional error term of the order $\frac{1}{n}$ to the certificate. These results, however, typically impose strong assumptions on the distributions (e.g. finite support or sub-Gaussianity) or not intended to be used optimally in the high-dimensional setting, for instance.

In addition to building a certificate on the loss under the data-generating distribution, the finite sample guarantees of the DRO estimator's out-of-sample performance are also established through the finite-sample upper bound of:
\begin{enumerate}
    \item $\EE_{\PP_{\star}}[\ell(\hat \theta_n, \xi)] - \inf_{\theta \in \Theta} \EE_{\PP_{\star}}[\ell(\theta, \xi)]$: in the context of applying shrinking radius $\delta_n$ when the out-of-sample environment is $\PP_{\star}$ (\citet[Corollary 5]{duchi2018variance}, \citet[Remark 1]{ gao2022finite}).
    \item $\sup_{\QQ \in {\mc B}_{\delta}(\PP_{\star})}\EE_{\QQ}[\ell(\hat \theta_{n,\delta}, \xi)] - \inf_{\theta \in \Theta}\sup_{\QQ \in {\mc B}_{\delta}(\PP_{\star})} \EE_{\QQ}[\ell(\theta, \xi)] $: in the context of applying fixed radius $\delta$ when the out-of-sample environment may have distributional shift from $\PP_{\star}$ (\citet[Theorem 3]{lee2018minimax}, \citet[Corollary 2]{duchi2021learning}).
\end{enumerate}

\subsubsection{Optimality of DRO Estimators}
In addition to its success in empirical studies, theoretical investigations into the DRO aim to establish a statistical optimality argument, further demonstrating its advantages. Several endeavors have provided potential avenues.

In a recent study, \citet{van2021data} justifies the optimality of DRO formulation with the help of the large deviation principles. The authors concentrate on a specific class of certificates, stipulating that the probability of underestimation of the expected loss under 
$\PP_{\star}$ decays exponentially fast as the sample size $n$ grows to infinity (similar to equation \eqref{eq:certificate}). The optimal certificate (in the sense of being the least conservative) is shown to be equivalent to a DRO formulation based on $\phi$-divergence (in particular corresponding to inverse KL). For those readers familiar with large deviations theory, the result may be natural given that it formally corresponds to an application of Sanov's theorem. This result is interesting because it provides an interpretation of DRO estimators as being optimal in some sense. However, the optimality criterion hinges on stipulating a large deviation gap in expected loss estimation (which traduces to a fixed uncertainty radius), which is too pessimistic a criterion as an inferential tool.

In another study on the DRO with fixed radius $\delta$ in the context of $\phi$-divergence-based uncertainty set, \cite{duchi2021learning} proves that the DRO estimator can achieve a sharp minimax bound for the loss 
\begin{align*}
    \sup_{\QQ \in {\mc B}_{\delta}(\PP_{\star})}\EE_{\QQ}[\ell(\hat \theta_{n,\delta}, \xi)] - \inf_{\theta \in \Theta}\sup_{\QQ \in {\mc B}_{\delta}(\PP_{\star})} \EE_{\QQ}[\ell(\theta, \xi)],
\end{align*}
where $\hat \theta_{n,\delta}$ is defined in \eqref{eq:DROfix}. This result demonstrates the ability of DRO estimator to learn with uniformly good performance. However, the loss function $\sup_{\QQ \in {\mc B}_{\delta}(\PP_{\star})}\EE_{\QQ}[\ell(\cdot, \xi)]$ for the DRO estimators to achieve optimal rate still appears too conservative in practice. Further, whether the DRO estimator can achieve a sharp minimax bound for other context such as optimal transport discrepancy is open.

\subsubsection{Statistical Inference of DRO} \label{subsect:confidence}
The uncertainty set on the distribution implemented in the DRO inspires new confidence regions of the statistics under the true data-generating distribution, such as
\begin{align*}
    \theta_{\star} = \argmin_{\theta \in \Theta} \EE_{\PP_{\star}}[\ell(\theta, \xi)]\quad \text{and} \quad
    \EE_{\PP_{\star}}[\ell(\theta_{\star}, \xi)] = \min_{\theta \in \Theta} \EE_{\PP_{\star}}[\ell(\theta, \xi)].
\end{align*}

For the confidence region of $\theta_{\star}$, the radius selection rule \eqref{eq:rwpi} suggests a suitable candidate
\begin{align*}
    \Lambda_{\delta_n} = \bigcup \left\{\theta(\QQ): \QQ \in {\mc B}_{\delta_n}(\hat \PP_n)\right\},
\end{align*}
where $\theta(\PP) = \mathop{\arg\min}_{\theta\in\Theta} \EE_{\PP}[\ell(\theta, \xi)]$. To select $\delta_n$, note that
\begin{align*}
    & \PP_{\star}\left(\theta(\PP_{\star}) \in \Lambda_{\delta_n}\right) \\
    =~& \PP_{\star}\left(\exists \QQ \in {\mc B}_{\delta_n}(\hat \PP_n): \theta(\PP_{\star}) \in \mathop{\arg\min}_{\theta\in\Theta}\EE_{\QQ}[\ell(\theta,\xi)] \right) \\
    =~& \PP_{\star}\left(\min_{\QQ: \theta(\PP_{\star}) \in \mathop{\arg\min}_{\theta\in\Theta}\EE_{\QQ}[\ell(\theta,\xi)]} \mathds{D}(\hat \PP_n, \QQ) \leq \delta_n \right). 
\end{align*}
Then the problem is reduced to compute the quantile of the projection distance
\begin{align*}
    \min_{\QQ: \theta(\PP_{\star}) \in \mathop{\arg\min}_{\theta\in\Theta}\EE_{\QQ}[\ell(\theta,\xi)]} \mathds{D}(\hat \PP_n, \QQ).
\end{align*}
For $\mathds{D} = \mathds{D}_{\phi}$, the quantile is related to Owen's theory of the empirical likelihood~\citep{owen2001empirical}. As a result, by \citet[Theorem 3.4]{owen2001empirical},
\begin{align*}
    n \times \min_{\QQ: \theta(\PP_{\star}) \in \mathop{\arg\min}_{\theta\in\Theta}\EE_{\QQ}[\ell(\theta,\xi)]} \mathds{D}_{\phi}(\hat \PP_n, \QQ) \overset{d}{\rightarrow} \frac{\phi''(1)}{2}\chi^{2}_{(q)},  
\end{align*}
where $\chi^{2}_{(q)}$ is the chi-square distribution with degree of freedom $q$ and $q$ depends on $\ell(\cdot, \cdot), \theta_{\star}, \PP_{\star}$. Therefore, given the confidence level $1-\alpha$ , $\delta_n$ is taken as $\frac{\bar \delta}{n}$, where $\bar \delta$ is the $(1-\alpha)$-quantile of the distribution $\frac{\phi''(1)}{2}\chi^{2}_{(q)}$.

\begin{figure}[ht]
    \centering
    \includegraphics[width=0.6\textwidth]{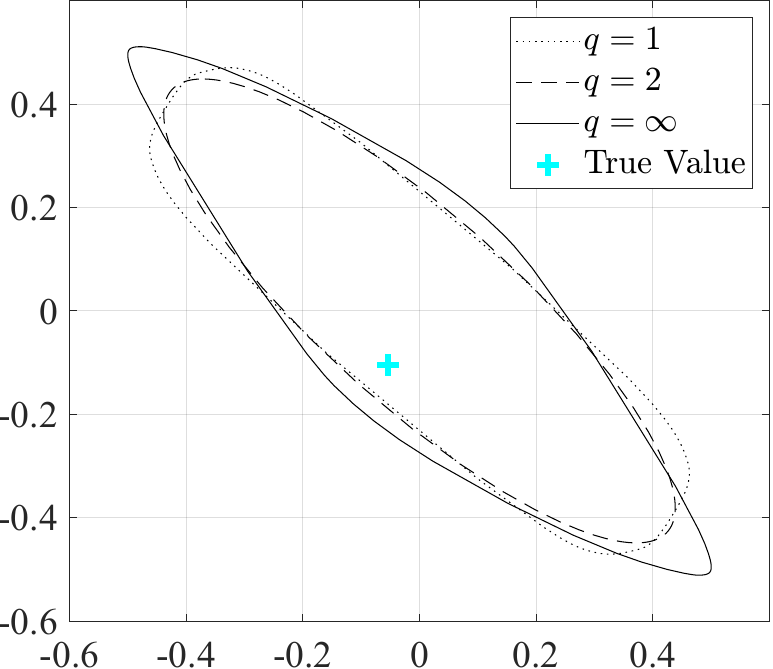}
    \caption{Confidence regions induced by optimal transport discrepancy-based DRO using different Euclidean distances as cost functions.}
    \label{fig:CRs}
\end{figure}

For $\mathds{D} = \mathds{D}_{c}$, \citet{blanchet2019robust} provides the quantile when the cost function is defined by the Euclidean distances. For cost functions of the form $c(\xi, \xi') = \|\xi - \xi'\|_{q}^2$, similar to the context of the $\phi$-divergence,  $\delta_n$ is taken as $\frac{\bar \delta}{n}$, where $\bar \delta$ depends on the quantile of limiting distribution of the projection distance given a confidence level. Specifically, \citet{blanchet2019robust} considers cost functions $c(\xi, \xi') = \|\xi - \xi'\|_{q}^2$ for various values of $q$, implying different geometry used when constructing the confidence region. Figure~\ref{fig:CRs} shows the confidence regions of the linear regression estimator computed in \citet[Section 4.2]{blanchet2019robust}.

For the confidence interval of $\EE_{\PP_{\star}}[\ell(\theta_{\star}, \xi)]$, it is straightforward to use the optimal value of the corresponding data-driven DRO problem as an upper bound. To get a lower bound, we consider an alternative problem with the $\sup$ inside of DRO replaced by $\inf$ \citep{lam2017empirical}. Let 
\begin{align*}
    & u_n = \min_{\theta\in\Theta}\sup_{\QQ \in {\mc B}_{\delta_n}(\hat \PP_{n})} \EE_{\QQ}[\ell(\theta, \xi)], \\
    & l_n = \min_{\theta\in\Theta}\inf_{\QQ \in {\mc B}_{\delta_n}(\hat \PP_{n})} \EE_{\QQ}[\ell(\theta, \xi)],
\end{align*}
with $\delta_n$ chosen similar to \eqref{eq:rwpi} for a specified threshold $\alpha \in (0,1)$, then
\begin{align*}
    \liminf_{n\rightarrow\infty}\PP_{\star}\left(\min_{\theta\in\Theta}\EE_{\PP_{\star}}[\ell(\theta,\xi)] \in [l_n, u_n] \right) \geq 1- \alpha.
\end{align*}
If additional smoothness of $\ell$ is assumed, then the asymptotic coverage can be exactly $1-\alpha$ with a smaller $\delta_n$~\citep[Theorem 3]{duchi2021statistics}.

\subsection{Tractability of DRO}\label{sec:tractable}
In this subsection, we discuss the tractability of the DRO formulations. 
To be more specific, the computation complexity of DRO problems depends on both the loss function and the uncertainty set, and thus we list some existing algorithms case by case. 
It's worth noting that the development of fast algorithms for solving different DRO models is still in its initial stages.

Starting from the moment-based DRO~\citep{delage2010distributionally}, the authors have paid attention to the tractable algorithms to solve the DRO problems, and specifically for the moment-based DRO problem, a reformulation as semidefinite programming is available. For the DRO based on optimal transport discrepancy, if the loss function is piecewise concave with respect to the random input, then the DRO can be reformulated as a conic programming and can be solved using general nonlinear optimization solver, e.g., MOSEK or Gurobi; if the loss function for linear decision rule is considered, \citet{li2019first, li2020fast, blanchet2022optimal} propose some fast first-order methods; if the loss function is parameterized as a neural network, some certificate is tractable as shown in~\citet{sinha2018certifying}. For the $\phi$-divergence-based DRO and convex loss function with respect to the parameter, \citet{levy2020large} proposes the mini-batch gradient descent method with complexity that is independent of sample size and dimension, and optimally dependent on the uncertainty set size.

Some evidence implies that the DRO can even bring benefits to the computation. For example, when $\ell(\theta,\xi)$ is strongly convex in $\theta$, for a suitable selection of the radius $\delta$, it is shown that the corresponding DRO problem is also strongly convex in $\theta$~\citep[Theorem 4]{blanchet2022optimal}, while the original non-DRO problem (i.e. $\delta=0$) may not be strongly convex, for example, in the high dimensional regression problem.

\subsection{DRO in the Bayesian Framework}\label{sec:Bayes}

A famous quote, often attributed to the statistician George Box~\citep{box1976science} indicates that ``every model is wrong, but some are useful.'' Our previous discussion of the DRO framework shows its flexibility in addressing issues related to model misspecification, emphasizing the deviation of out-of-sample distribution from the samples' empirical distribution employed by data-driven methods. However, in a model-driven setting, the issue of misspecification may be even more severe. As an extension to enhancing the robustness of the data-driven decision-making cycle, this subsection illustrates the application of DRO to mitigate the impact of model misspecification within the framework of Bayesian statistics. We emphasize that this area is significantly less investigated, but still, there is a substantial and growing literature that studies this setting.  

In the Bayesian framework, the sensitivity of the model performance to the perturbations of the prior and likelihood is an important topic. While there is a rich literature in Bayesian statistics to model robustness, the vast majority of which are related to the specification of prior. Interested readers are referred to another review paper~\citet{watson2016approximate}. Actually, the robustness with respect to the specification of likelihood is no less important than the prior. 
However, modeling and hedging against the distributional misspecification of the likelihood is less studied in the Bayesian framework, possibly because robust formulations of the likelihood naturally lead to infinite dimensional (i.e. non-parametric) formulations, and this results in significantly higher complexity. In contrast, the DRO framework routinely manages non-parametric distributional uncertainty sets, making it conducive for integration into the Bayesian framework. This strategy will be illustrated through examples in the following discussion.   

In this subsection, we assume that $\eta \in \R^{p}$ denotes a prior parameter, $\xi\in \R^{d}$ denotes the observed sample following a distribution parameterized by $\eta$, and $\hat \theta(\cdot)\in \Phi$ denotes a decision policy to be taken based on the full Bayesian model. For example, in an investment problem, $\eta$ denotes the (unknown) vector of mean of the distribution of returns, the returns are denoted as $\xi$, and $\hat \theta(\xi)$ is the vector of portfolio allocations based on the observed returns. The decision maker is interested in minimizing the risk, which can be taken as the negative of a utility. 

Precisely, assume that $\ell: \Phi \times \Xi \times \mathcal{H}\rightarrow \R$ is the loss function, $\QQ_{\xi|\eta}$ is the likelihood and $\QQ_{\eta}$ is the prior, we consider the stochastic optimization problem 
\begin{align*}
    \min_{\hat \theta(\cdot) \in \Phi}~\EE_{\QQ}\left[\ell(\hat \theta(\cdot), \xi, \eta)\right] = \EE_{\QQ_{\eta}}\left[ \EE_{\QQ_{\xi | \eta}}\left[\ell(\hat \theta(\cdot), \xi, \eta)\right]\right].
\end{align*}
Given $\xi_0 \in \Xi$, we also consider the problem with a slightly different objective 
\begin{align*}
    \min_{\hat \theta(\cdot) \in \Phi}~\EE_{\QQ}\left[\ell(\hat \theta(\cdot), \xi, \eta) | \xi = \xi_0\right].
\end{align*}
For example, when considering the Bayesian minimum mean square estimation problem, we can set $\ell(\hat \theta(\cdot), \xi, \eta) = \|\eta - \hat \theta(\xi)\|_2^2$.

We now review recent work on the DRO frameworks built on the aforementioned objectives and applied to typical problems in robust Bayesian inference \citep{levy2012robust, zorzi2016robust, shafieezadeh2018wasserstein,nguyen2020robust,zhang2022wasserstein, nguyen2023bridging, lotidis2023wasserstein}. The DRO framework introduced in these studies includes the uncertainty on both prior and likelihood, which extends some of the classic studies on robust prior involving both prior and likelihood misspecification.

\begin{enumerate}
    \item[(i).] \citet{levy2012robust, zorzi2016robust, shafieezadeh2018wasserstein,lotidis2023wasserstein} consider the Kalman filtering: for $i=1,\dots,T$,
    \begin{align*}
            \eta_i = D_i \eta_{i-1} + v_i,\quad \xi_i = B_i \eta_i + u_i,
    \end{align*}
    where $v = (v_1, \dots, v_T)$ denotes the innovation process and $u = (u_1, \dots, u_T)$ denotes the noise process. Let $\xi = (\xi_i, 1 \leq i\leq T)$ and $\eta = (\eta_i, 1\leq i \leq T)$, the loss function is set to be
    \begin{align*}
        \ell(\hat \theta(\cdot), \xi, \eta) = \sum_{i=1}^{T} \left\| \eta_i - \hat \theta_i(\xi_1, \dots, \xi_i) \right\|^2.
    \end{align*}
    where $\hat \theta(\cdot) = (\hat \theta_i(\cdot), 1\leq i\leq T)$ are the linear functionals to be learned. In the corresponding DRO formulation, the authors consider an uncertainty set on the joint distribution of $(\eta_1,\dots,\eta_T)$ and $(\eps_1, \dots, \eps_T)$, that is, given $\xi = \xi_{0}$,
    \begin{align*}
        \min_{\hat \theta(\cdot) \in \Phi} \sup_{\QQ \in {\mc B}} \EE_{\QQ}\left[\ell(\hat \theta(\cdot), \xi, \eta) | \xi = \xi_{0}\right],
    \end{align*}
    where $\mc B$ is an uncertainty set that is defined by a neighborhood ball centered at some nominal distribution, measured by the $\phi$-divergence \citep{levy2012robust, zorzi2016robust} or the optimal transport discrepancy \citep{shafieezadeh2018wasserstein,lotidis2023wasserstein}. In the case of $\phi$-divergence-based uncertainty set, as it turns out, the decision rule often remains unchanged compared to the non-DRO case, but the mean squared error naturally increases. In contrast, in the case of optimal transport-based uncertainty set, the decision rule undergoes a change. In both cases, if the baseline distribution is Gaussian, the worst-case distribution remains Gaussian and, therefore, affine decision rules are optimal when minimizing over the class arbitrary prediction functions. This result is also shown in \citet{nguyen2023bridging}, where the authors consider the general Bayesian minimum mean square error estimation problem with the optimal transport-based uncertainty set.

    \item[(ii).] \citet{nguyen2020robust} considers the Bayesian classification problem with the classification error to be 
    \begin{align*}
        \ell(\hat \theta(\cdot), \xi, \eta) = \hat \theta(\xi) \mathbb{I}_{\eta = 0} + (1-\hat \theta(\xi)) \mathbb{I}_{\eta=1},
    \end{align*}
    where $\eta\in\{0,1\}$ denotes the unobserved label, $\xi \in \Xi$ is the observed sample to be classified, and $\hat \theta: \Xi \rightarrow \{0,1\}$ is the (randomized) classifier to be learned. The authors consider the DRO formulation of an objective, that given $\xi = \xi_0$,
    \begin{align*}
        \min_{\hat \theta(\cdot) \in \Phi} \sup_{\QQ \in {\mc B}} \EE_{\QQ}\left[\ell(\hat \theta(\cdot), \xi, \eta) | \xi = \xi_0\right].
    \end{align*}
    Specifically, the uncertainty set $\mc B$ is built on the joint distribution of prior and likelihood and is set to be
    \begin{align*}
        \mc B = 
        \left\{
        \QQ = \pi_0 \QQ_{\xi| \eta = 0} + \pi_1 \QQ_{\xi|\eta = 1}: 
        \begin{array}{cc}
            \QQ_{\xi| \eta = 0} \in \mc P, &  \mathds{D}_{\phi}(\QQ_{\xi| \eta = 0}, \QQ_0) \leq \delta_0\\
            \QQ_{\xi| \eta = 1} \in \mc P, &  \mathds{D}_{\phi}(\QQ_{\xi| \eta = 1}, \QQ_1) \leq \delta_1 
        \end{array}
        \right\},
    \end{align*}
    where $\delta = (\delta_0, \delta_1)$, $(\pi_0, \pi_1)$ is a specified prior weight on $\eta$, $(\QQ_0, \QQ_1)$ are two nominal distributions, and $\mc P$ is a specified parametric family.

    \item[(iii).] \citet{zhang2022wasserstein} considers Gaussian process regression and linear inverse problems. This is one of the few DRO results that involve non-parametric decision rules, non-parametric distributional uncertainty, and an infinite dimensional outcome space. In particular, the formulation is as follows,
    \begin{align*}
        \xi_i = \eta(t_i) + \eps_i,\qquad i = 1,\dots,n.
    \end{align*}
    where $\xi = (\xi_i, 1\leq i \leq n)$ are observed values in $\R$, $\eta$ is the parameter of interest, which is modeled as a real-valued random process (endowed with prior) with continuous sample paths, $(t_i, 1\leq i \leq n)$ are some specified evaluation points in a domain $\mc D \subset \R^d$, and $(\eps_i, 1\leq i\leq n)$ denote the observational noise. The loss function is
    \begin{align*}
        \ell(\hat \theta(\cdot), \xi, \eta) = \left\| \eta - \hat \theta(\xi_1, \dots, \xi_n)\right\|^2_{\mc L^2(\mc D)},
    \end{align*}
    where the estimator $\hat \theta: \R^{n} \rightarrow \mc L^2(\mc D)$ maps the observed values $(\xi_i, 1\leq i\leq n)$ to a continuous path on the domain $\mc D$. In the corresponding DRO formulation, the authors consider the uncertainty set on the joint distribution of the sample path $\eta$ (prior) and $\xi$ (likelihood) using the optimal transport discrepancy. In particular, they consider the reproducing kernel Hilbert space (RKHS) to characterize the neighborhood of a sample path of $\eta$ and further the neighborhood of the nominal distribution of $\eta$. Remarkably, the authors show that if the ground transportation cost function $c$ is a squared Hilbert norm and the nomial distribution at the center of the uncertainty set is a Gaussian process, then the worst-case distribution is also a Gaussian process, and therefore the optimal prediction function is affine in the observations. Moreover, the authors show that a Nash equilibrium exists and it is unique for sufficiently small uncertainty budgets.
\end{enumerate}

\section{Robust Statistics}\label{sec:robuststat}

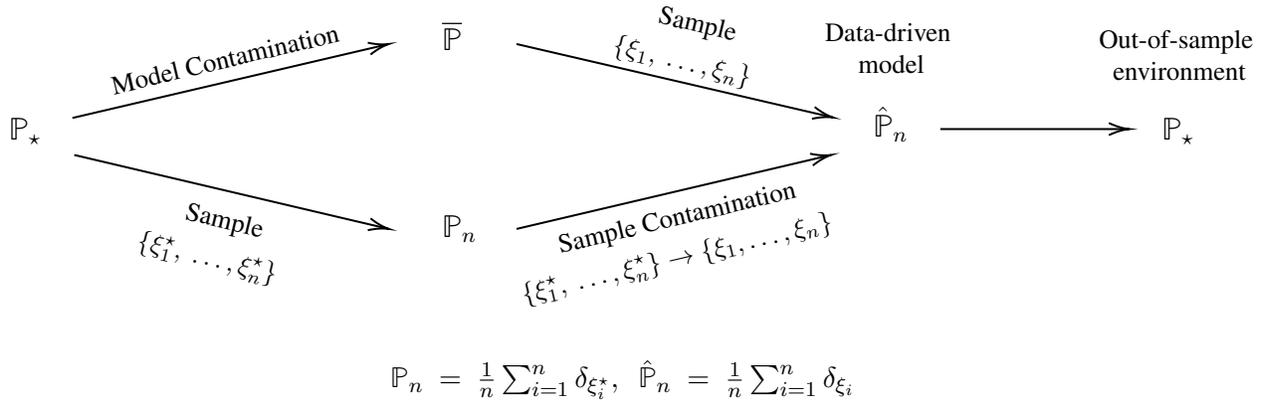
\begin{figure}[ht]
    \centering

\tikzset{every picture/.style={line width=0.75pt}} 

\begin{tikzpicture}[x=0.7pt,y=0.7pt,yscale=-1,xscale=1]

\draw    (42.86,131.14) -- (211.91,91.6) ;
\draw [shift={(213.86,91.14)}, rotate = 166.83] [color={rgb, 255:red, 0; green, 0; blue, 0 }  ][line width=0.75]    (10.93,-3.29) .. controls (6.95,-1.4) and (3.31,-0.3) .. (0,0) .. controls (3.31,0.3) and (6.95,1.4) .. (10.93,3.29)   ;
\draw    (282.86,191.14) -- (450.91,151.6) ;
\draw [shift={(452.86,151.14)}, rotate = 166.76] [color={rgb, 255:red, 0; green, 0; blue, 0 }  ][line width=0.75]    (10.93,-3.29) .. controls (6.95,-1.4) and (3.31,-0.3) .. (0,0) .. controls (3.31,0.3) and (6.95,1.4) .. (10.93,3.29)   ;
\draw    (43.86,151.14) -- (211.91,190.68) ;
\draw [shift={(213.86,191.14)}, rotate = 193.24] [color={rgb, 255:red, 0; green, 0; blue, 0 }  ][line width=0.75]    (10.93,-3.29) .. controls (6.95,-1.4) and (3.31,-0.3) .. (0,0) .. controls (3.31,0.3) and (6.95,1.4) .. (10.93,3.29)   ;
\draw    (282.86,91.14) -- (451.91,130.4) ;
\draw [shift={(453.86,130.86)}, rotate = 193.07] [color={rgb, 255:red, 0; green, 0; blue, 0 }  ][line width=0.75]    (10.93,-3.29) .. controls (6.95,-1.4) and (3.31,-0.3) .. (0,0) .. controls (3.31,0.3) and (6.95,1.4) .. (10.93,3.29)   ;
\draw    (511.86,137.14) -- (611.86,137.14) ;
\draw [shift={(613.86,137.14)}, rotate = 180] [color={rgb, 255:red, 0; green, 0; blue, 0 }  ][line width=0.75]    (10.93,-3.29) .. controls (6.95,-1.4) and (3.31,-0.3) .. (0,0) .. controls (3.31,0.3) and (6.95,1.4) .. (10.93,3.29)   ;

\draw (7,129.4) node [anchor=north west][inner sep=0.75pt]  [font=\large]  {$\mathbb{P}_{\star }$};
\draw (241,77.4) node [anchor=north west][inner sep=0.75pt]  [font=\large]  {$\overline{\mathbb{P}}$};
\draw (474,123.4) node [anchor=north west][inner sep=0.75pt]  [font=\large]  {$\hat{\mathbb{P}}_{n}$};
\draw (630,129.4) node [anchor=north west][inner sep=0.75pt]  [font=\large]  {$\mathbb{P}_{\star }$};
\draw (61.2,107.57) node [anchor=north west][inner sep=0.75pt]  [font=\small,rotate=-346.7] [align=left] {Model Contamination};
\draw (277.69,199.26) node [anchor=north west][inner sep=0.75pt]  [font=\small,rotate=-346.7] [align=left] {\begin{minipage}[lt]{127.59pt}\setlength\topsep{0pt}
\begin{center}
Sample Contamination
\end{center}
$\displaystyle \{\xi^{\star}_{1} ,\ \dotsc ,\xi^{\star}_{n}\}\rightarrow \{\xi _{1} ,\dotsc ,\xi _{n}\}$
\end{minipage}};
\draw (83.57,169.21) node [anchor=north west][inner sep=0.75pt]  [font=\small,rotate=-13.7] [align=left] {\begin{minipage}[lt]{60.14pt}\setlength\topsep{0pt}
\begin{center}
Sample 
\end{center}
$\displaystyle \{\xi^{\star}_{1} ,\ \dotsc ,\xi^{\star}_{n}\}$
\end{minipage}};
\draw (340.68,63.18) node [anchor=north west][inner sep=0.75pt]  [font=\small,rotate=-13.69] [align=left] {\begin{minipage}[lt]{58.07pt}\setlength\topsep{0pt}
\begin{center}
Sample 
\end{center}
$\displaystyle \{\xi _{1} ,\ \dotsc ,\xi _{n}\}$
\end{minipage}};
\draw (241,182.4) node [anchor=north west][inner sep=0.75pt]  [font=\large]  {$\mathbb{P}_{n}$};
\draw (448,78) node [anchor=north west][inner sep=0.75pt]  [font=\small] [align=left] {\begin{minipage}[lt]{50.14pt}\setlength\topsep{0pt}
Data-driven
\begin{center}
model
\end{center}

\end{minipage}};
\draw (596,82) node [anchor=north west][inner sep=0.75pt]  [font=\small] [align=left] {\begin{minipage}[lt]{60.84pt}\setlength\topsep{0pt}
Out-of-sample
\begin{center}
environment
\end{center}

\end{minipage}};
\draw (213,260.4) node [anchor=north west][inner sep=0.75pt]    {$\mathbb{P}_{n} \ =\ \frac{1}{n}\sum _{i=1}^{n} \delta _{{\xi }^{\star}_{i}} ,\ \ \hat{\mathbb{P}}_{n} \ =\ \frac{1}{n}\sum _{i=1}^{n} \delta _{\xi _{i}}$};

\end{tikzpicture}

    \caption{Decision Making Cycle in the Context of Robust Statistics}
    \label{fig:RobustStatistics}
\end{figure}


In the context of robust statistics, robust estimators act after the occurrence of pre-decision distributional contamination. For example, in scenarios involving noisy observations, samples are i.i.d.~generated from a contaminated data-generating distribution. In another agnostic setting, a more potent adversary is assumed to contaminate the samples after they are generated from the true data-generating distribution. To distinguish between these two types of pre-decision contamination, we refer to Figure~\ref{fig:RobustStatistics}, which, in contrast to Figure~\ref{fig:decisioncycle}, excludes post-decision contamination and focuses on estimating the true data-generating distribution. A more detailed discussion of types of pre-decision contamination will be presented in Section~\ref{sec:typeofcontamination}.

To clarify our notations in this section, we use the following symbols: $\overline \PP$ for the contaminated data-generating distribution, ${\PP}_n$ for the empirical distribution composed of uncontaminated samples, and $\hat {\PP}_n$ for the empirical distribution composed of contaminated samples (also representing the observed samples, consistent with the notation used in the discussion of DRO). We will use $\hat {\PP}_n$ interchangeably to denote both the contaminated samples and their empirical distribution, and ${\PP}_n$ interchangeably to denote both the uncontaminated samples and their empirical distribution.

In the subsequent sections, we embark on a comprehensive review of the field of robust statistics. Following this, we draw comparisons between robust statistics and DRO. For ease of exposition, our focus will be on the fundamental task of robust estimation for the location parameter, specifically robust mean estimation, serving as a running example in our review.

\subsection{Literature Review}
The field of robust statistics emerged from statistical procedures tailored to confront two practical challenges: the detection and rejection of ``outliers'', and the analysis of data when the underlying distribution deviates from normality. Historically, there was ambiguity associated with the concept of robustness~\citep{huber1972review}. The term ``robustness'' appeared to have first been introduced by Box~\citep{box1953nonnormality}, who used the term to refer to the property of a procedure insensitive to departures from ideal assumptions~\citep{BOX1979robustness}. Tukey also pioneered the recognition of sensitivity of some conventional
 statistical procedures to minor deviations from the assumptions~\citep{tukey1959survey,Tukey1962future}, and developed a series of work in nonparametric methods and rank-based procedures~\citep{scheffe1944formula,scheffe1945nonpara}. The early foundations of robust estimation were further developed by Huber~\citep{huber1964robust,Huber1968confidence} and Hampel~\citep{hampel1968contributions,hampel1971general}, among others. To specify the setup of robust estimation considered here, we again assume that $\xi$ is a random vector in the space $\Xi$ that follows a data-generating distribution $\PP_{\star}$ (the canonical example being a normal distribution), and consider $\theta\in\Theta$ as the parameter of the model to be learned. In particular, we consider $\theta=\theta(\PP_{\star})$ coming from a statistical functional $\theta(\cdot): \mc P(\Xi) \rightarrow \Theta$. In the framework of $M$-estimators~\citep{vaart1998asymptotic}, the functional $\theta(\PP_{\star})$ would be a minimizer of the criterion function

 \begin{align}\label{eq:ERM2}
    \min_{\theta \in \Theta} \EE_{\PP_\star}[\ell(\theta, \xi)] = \int_{\R^d} \ell(\theta, \xi) \diff \PP_\star (\xi),
\end{align}
where $\ell(\theta,\xi)$ is a measurable extended real-valued function. For example, to obtain the mean parameter it is typical to consider the squared loss $\ell(\theta,\xi)= \|\theta-\xi\|_2^2$. Given observations $\xi_1,\ldots,\xi_n$ each of which follows marginal law $\PP_{\star}$, and are assumed to be independent, the $M$-estimator is the solution to the empirical criterion function
\begin{align}\label{eq:erm2}
    \min_{\theta \in \Theta} \EE_{\hat \PP_n}[\ell(\theta, \xi)] = \frac{1}{n} \sum_{i=1}^{n} \ell(\theta, \xi_i),
\end{align}
where $\hat \PP_n$ denotes the empirical measure $\frac{1}{n}\sum_{i=1}^{n}\delta_{\xi_i}$. In the example of mean estimation, the solution turns out to be the sample mean. In practice, the in-sample empirical measure $\hat \PP_n$ may deviate from the ideal assumptions imposed, consequently compromising the accuracy of naive $M$-estimators. Regarding sample mean, the presence of even a single outlier within the data points can substantially undermine the statistical performance of this estimator. There are several different reasons for the in-sample empirical measure $\hat \PP_n$ to deviate from the ideal assumptions. To be precise, now we review the type of deviations typically considered in the literature. These mechanisms of deviations are also called contamination models (or corruption models).

\subsubsection{Types of Contamination Models}\label{sec:typeofcontamination}
Conceptually, we consider the existence of an adversary, who generates the in-sample data set in an adversarial fashion. As shown in Figure~\ref{fig:RobustStatistics}, the adversary can perturb the true data-generating distribution $\PP_{\star}$ to a contaminated data-generating distribution $\overline{\PP}$, after which the samples $\hat \PP_n$ are drawn i.i.d.~from $\overline{\PP}$. A stronger form of contamination is that the adversary can perturb the in-sample data set \textit{adaptively}, which can contaminate the samples $\PP_n$ that are directly generated from the uncontaminated data-generating distribution $\PP_{\star}$, resulting in the contaminated samples $\hat \PP_n$. Specifically, we mention

 \textrm{(i)} \textit{Huber's $\epsilon$-contamination model}: In a seminal paper~\citet{huber1964robust}, Huber considered the contamination for the data-generating distribution as $\overline{\PP} = (1-\epsilon) \PP_{\star}+ \epsilon \mathbb{H}$. In his original paper $\PP_{\star}$ is a normal distribution, $\mathbb{H}$ is an unknown contamination distribution, and $\epsilon$ is a known constant representing the level of contamination. Hence, the adversary is allowed to add contamination to (but not subtract from) the population distribution that generates the samples.
 
\textrm{(ii)} \textit{Full-neighborhood contamination}: Instead of the restricted neighborhood in Huber's contamination model, the adversary perturbs the data-generating distribution in a full-neighborhood by the use of a statistical distance $\mathds{D}(\overline{\PP},\PP_{\star})\leq\epsilon$.  When $\mathds{D}$ is chosen as the total variation distance, this neighborhood is strictly larger than Huber's contamination. The choice of total variation distance is typically for the study of gross errors in the dataset~\citep{donoho1988automatic,zhu2022generalized}. The statistical distance can be modified to study other types of natural model errors, such as rounding errors. For instance, refer to~\citet{zhu2022generalized,liu2022robust} for the use of Wasserstein distance of order $1$.

 \textrm{(iii)} \textit{Adaptive contamination model}: A more powerful contamination works as follows. Once the samples ${\PP}_n$ are drawn~i.i.d. from $\PP_{\star}$, the adversary inspects the samples, and at their disposal, remove up to $\epsilon  n$ samples and replace them with arbitrary points, resulting in the contaminated in-sample data set $\hat{\PP}_n$~\citep{diakonikolas2019robust,zhu2022generalized}. This is equivalent to the constraint that $\mathds{D}({\PP}_n, \hat\PP_n)\leq \epsilon $, where $\mathds{D}$ is chosen as the total variation distance. To model the scenario where every sample can be slightly perturbed, ~\citet{zhu2022generalized,liu2022robust} also consider the use of Wasserstein distance of order $1$.

\subsubsection{Types of Robustness Criteria} 
The objective of robust statistics is to develop estimators that exhibit certain ``robustness'' properties in relation to the aforementioned forms of contamination. There are multiple criteria to quantify robustness. In this discussion, we examine some of the robustness criteria.

To restate our notations corresponding to the three contamination models previously described, we generically denote $\mathcal{A}_\epsilon(\PP_\star) = \{\PP: \mathds{D}(\PP, \PP_\star) \leq \eps\}$ to be a set of contamination models with probability metric $\mathds{D}$ to be further specified, $\overline{\PP}\in\mathcal{A}_\epsilon(\PP_\star)$ to represent the population distribution from which the contaminated samples $\hat \PP_n$ are drawn. This draw is~i.i.d for the first two types of contamination models aforementioned. For the adaptive contamination model, we denote $\hat{\PP}_n\in\mathcal{A}_\epsilon({\PP}_n)$ as the empirical distribution of samples modified from ${\PP}_n$, which is the empirical distribution of samples i.i.d. generated from $\PP_\star$. Further, we generically denote $\hat\theta=\hat\theta(\hat \PP_n)$ as the estimator of interest.

Now we start to review the robustness criteria, which include:

\textrm{(i)} \textit{Efficiency}: A small contamination level should cause a small degradation on the statistical performance. An intuitive formal criterion to minimize would be
    \begin{equation}\label{eq:errstability}
        err(\epsilon,\PP_\star) = \sup_{\overline{\PP}\in\mathcal{A}_\epsilon(\PP_\star)}\EE_{\overline{\PP}}\left[\EE_{\PP_\star}[\ell(\hat\theta(\hat \PP_n), \xi)]\right],
    \end{equation}
    where the outer expectation $\EE_{\overline{\PP}}$ averages the randomness over the draw of the contaminated samples $\hat \PP_n$ and the inner expectation $\EE_{\PP_\star}$  denotes averaging over the draw of $\xi$ following the true data-generating distribution $\PP_\star$. The corresponding criterion to minimize for the adaptive contamination model would be
    \[
    err(\epsilon,{\PP}_n) = \sup_{\hat \PP_n\in\mathcal{A}_\epsilon({\PP}_n)}\EE_{\PP_\star}[\ell(\hat\theta(\hat\PP_n), \xi)],
    \]
    where $\EE_{\PP_\star}$ denotes averaging over the draw of $\xi$ following the true data-generating distribution $\PP_\star$.
    
    Choosing the squared loss $\ell(\theta,\xi)= \|\theta-\xi\|_2^2$ and the mean functional $\theta(\PP_\star)= \EE_{\PP_\star}[\xi]$, the above criteria are equivalent to 
    \[
    err(\epsilon,\PP_\star) = \sup_{\bar{\PP}\in\mathcal{A}_\epsilon(\PP_\star)}\EE_{\bar\PP}\left[\left\|\hat\theta(\hat \PP_n)-\theta(\PP_\star)\right\|_2^2\right],
    \]
    and
    \[
    err(\epsilon,{\PP}_n) = \sup_{\hat \PP_n\in\mathcal{A}_\epsilon({\PP}_n)}\left\|\hat\theta(\hat \PP_n)-\theta(\PP_\star)\right\|_2^2,
    \]
    respectively.
     For technical reasons, instead of the expected risk in~\eqref{eq:errstability}, prior works focused on establishing high probability bounds for the quantity 
    \[
    \EE_{\PP_\star}[\ell(\hat\theta(\hat \PP_n), \xi)],
    \]
where besides assumptions on the contamination models, usually some assumptions on $\PP_\star$ are imposed.
For instance, the class of normal or elliptical distributions was considered in~\citet{chen2018robustcov,chao2018robust, gao2020robustregression}, and the class of distributions having subgaussian tails or bounded moments was studied in~\citet{diakonikolas2017being, steinhardt2017certified,steinhardt2018resilience,diakonikolas2018learning,diakonikolas2019sever, Bateni2019ConfidenceRA,liu2020high,depersin2022robust}.

    \textrm{(ii)} \textit{Break-down point}: The notion of break-down point was introduced by Hampel~\citep{hampel1968contributions,hampel1971general}, and later developed by Donoho and Huber~\citep{donoho1983notion,huber2004robust} (among others) to quantify the influence of outliers on a given estimator $\hat\theta$. The finite-sample break-down point, following~\citet{donoho1983notion}, can be formulated as 
    \[
    \min\{\epsilon: \|\hat\theta({\PP}_n) - \hat\theta(\hat \PP_n)\|_2=\infty,~~\hat \PP_n\in\mathcal{A}_\epsilon({\PP}_n)\},
    \]
    the smallest fraction of corruption to the samples that causes the estimator to be arbitrarily bad. 
    For example, the break-down point for the sample mean is easy to be seen as $\frac{1}{n}$. A variant of the break-down point in the population sense can be similarly defined as 
    \[
     \min\{\epsilon: err(\epsilon,\PP_{\star})=\infty\},
    \]
    the smallest $\epsilon$ that causes the expected risk in~\eqref{eq:errstability} to be arbitrarily large.

Simultaneously achieving efficiency in terms of statistical convergence rate and a high break-down point is an on-going area of research. In~\citet{chen2018robustcov}, the authors propose a population variant
of the breakdown point which they term as ``$\delta$-breakdown point'', and show that
for a given estimator that has convergence rate $\delta$ under the Huber's $\epsilon$-contamination model, its $\delta$-breakdown point is at least $\epsilon$. This suggests that efficiency under Huber’s $\epsilon$-contamination
model may be a more general notion of robustness than the breakdown point.

\subsubsection{Early Work on Robust Mean Estimation}
In the seminal paper~\citet{huber1964robust}, for one-dimensional estimation, Huber considered an $M$-estimator by replacing the squared loss $\ell(\theta,\xi) = (\theta-\xi)^2$ with the loss function
\[
\ell(\theta,\xi) = \rho(\theta-\xi),
\]
where
\[
\rho(t) = \begin{cases}
    \frac{1}{2}t^2 & \text{for }|t|<k\\
    k|t|-\frac{1}{2}k^2 &\text{for }|t|\geq k,
\end{cases}
\]
and $k$ is related to $\epsilon$ by a non-linear equation. 
Huber demonstrated the optimality of this $M$-estimator among all translation-invariant estimators in terms of robustness. However, the measure of robustness considered therein is defined as the worst-case (suprema) of the asymptotic variance of the estimator over Huber's $\epsilon$-contamination model $\overline{\PP} = (1-\epsilon) \PP_{\star}+ \epsilon \mathbb{H}$, with normal $\PP_\star$ and symmetric $ \mathbb{H}$.

Tukey's median~\citep{Tukey1975Mathematics} is a robust mean estimator that can be defined in any dimension $d$. First, Tukey's depth function of any $\eta\in\R^d$ with respect to any distribution $\PP$ on $\R^d$ is defined as
\[
\mc D(\eta,\PP) = \inf_{u\in \mc S^{d-1}}\PP(u^\top \xi\leq u^\top \eta),\quad \xi\sim\PP,
\]
where $\mc S^{d-1}$ is the $d$-dimensional unit sphere in $\R^d$. Tukey’s median is defined to be
the deepest point with respect to the empirical distribution $\hat \PP_n$
\[
\hat\theta(\hat\PP_n) = \arg\max_\eta \mc D(\eta, \hat\PP_n).
\]
The convergence rate of Tukey’s median under Huber's $\epsilon$-contamination model is $\|\hat\theta(\hat \PP_n)-\theta(\PP_\star)\|^2_2\lesssim O\left(\frac{d}{n}\vee \epsilon^2\right)$, and such a finite-sample rate is also optimal in a minimax sense, as shown in~\citet{chen2018robustcov}. Unfortunately, it is well-known that Tukey's median is NP-hard to compute in
high dimensions~\citep{JOHNSON1978densest,bernholt2006robust}. A natural alternative is the componentwise median. Despite having a high break-down point~\citep{donoho1992breakdown}, this estimator suffers from a convergence rate $O\left(d\left(\frac{1}{n}\vee \epsilon^2
\right)\right)$, which is inferior in high dimensions~\citep{chen2018robustcov}.

\subsubsection{Recent Work on Computationally Efficient High-Dimensional Robust Mean Estimation}
Until recently, computationally efficient methods capable of achieving statistically optimal convergence rates were elusive. Naive polynomial time approaches typically lead to an error rate of $O(d\epsilon^2)$ (assuming enough number of samples), which scales linearly with the dimension $d$.  The recent works~\citep{diakonikolas2017being,diakonikolas2019robust} obtained the first dimension-independent error rate for computationally efficient robust mean estimation of an  isotropic normal distribution under the adaptive contamination model, and simultaneously,~\citet{lai2016agnostic} obtained an error rate that scales with $O(\log(d))$ under a weaker bounded fourth moment assumption. To obtain these dimension independent errors, fairly strong assumptions on the uncorrupted sample ${\PP}_n$ are imposed.
\begin{definition}[Stability~{\citep[Definition~2.1]{diakonikolas2023algorithmic}}]\label{def:stability}
    Fix $0 < \epsilon < 1/2$ and $\delta \geq \epsilon$. A finite set $S \subseteq \mathbb{R}^d$ is $(\epsilon, \delta)$-stable with respect to a vector $\mu$ if for every unit vector $v \in \mc S^{d-1}$ and every $S' \subseteq S$ with $|S'| \geq (1 - \varepsilon)|S|$, the following conditions hold:
\begin{enumerate}
\item $\left| \frac{1}{|S'|} \sum_{\xi \in S'} v \cdot (\xi - \mu) \right| \leq \delta$, and
\item $\left| \frac{1}{|S'|} \sum_{\xi \in S'} (v \cdot (\xi - \mu))^2 - 1 \right| \leq \frac{\delta^2}{\epsilon}$.
\end{enumerate}
Similarly, a distribution $\PP$ on $\mathbb{R}^d$ is $(\epsilon, \delta)$-stable with respect to a vector $\mu$ if for every unit vector $v \in \mc S^{d-1}$ and distribution $\bar\PP$ obtained from $\PP$ by conditioning on an event of probability at least $1-\epsilon$, the following conditions hold:
\begin{enumerate}
\item $\mathbb{E}_{\bar\PP}\left[|v \cdot (\xi - \mu)|\right] \leq \delta$, and
\item $\left|\mathbb{E}_{\bar\PP}\left[(v \cdot (\xi - \mu))^2\right] - 1\right| \leq \frac{\delta^2}{\varepsilon}$.
\end{enumerate}
\end{definition}
In simple terms, these conditions imply that the removal of any $\epsilon$-fraction of the sample points will not change the mean by more than $\delta$ nor the variance in any direction by more than $\frac{\delta^2}{\epsilon}$. It can be shown that these conditions are satisfied by many distributions (and with high probability their empirical samples) with appropriate concentration properties (e.g., normal distributions). The first condition in Definition~\ref{def:stability} is also studied under the name of \textit{resilience} by~\citet{steinhardt2018resilience}.

Under these stability conditions,~\citet{diakonikolas2019robust} proposed an iterative greedy algorithm aimed at purifying a dataset by progressively removing corrupted samples. Starting with an initial data set $S$ that includes both corrupted and uncorrupted samples, the algorithm in each iteration either calculates the sample mean of the current set of samples or employs a filter to refine $S$
  into a subset $S^{'}$
  that is substantially closer to the uncontaminated set.

Another approach proposed in~\citet{diakonikolas2019robust} is a convex programming based algorithm. Weights \( w_i \) are computed for each sample \( \xi_i \), so the weighted empirical average \( \hat{\mu}_{\boldsymbol{\omega}} = \sum_{i=1}^N \omega_i \xi_i \) approximates the true mean \( \mu \). These weights are constrained within a convex set \( C_\delta \), defined as:
\[
C_\delta = \left\{ \boldsymbol{\omega}: \begin{array}{l}
\sum_{i=1}^n \omega_i = 1,~0 \leq \omega_i \leq \frac{1}{(1 - \varepsilon)n}~\forall i \\
\left\| \sum_{i=1}^n \omega_i (\xi_i - \mu)(\xi_i - \mu)^\top - I \right\|_2 \leq \delta 
\end{array}\right\}.
\]
Since $\mu$ is unknown, the algorithm substitutes $\mu$ by $\hat\mu_{\boldsymbol{\omega}}$. Using spectral techniques to approximate a separation oracle for \( C_\delta \), the algorithm is shown to achieve computational efficiency in estimating \( \mu \). 

Since the works of~\citet{lai2016agnostic,diakonikolas2017being,diakonikolas2019robust}, the field has seen a proliferation of research. For a comprehensive overview, one can refer to the survey by~\citet{diakonikolas2019recent}. In a different thread of works,~\citet{chao2018robust}  established a connection
between generative adversarial networks~\citep{goodfellow2014generative} and classical depth-based robust estimators
(e.g. Tukey’s median), leading them to study robust
mean estimators using GANs and establish minimax
optimal error bounds. This connection also enables for the computation of robust estimators utilizing techniques originally developed for training GANs. The follow-up works extended the technique for $f$-GAN~\citep{wu2020minimax} and Wasserstein-GAN~\citep{liu2022robust}, under various contamination models.

\subsubsection{Information-Theoretical Lower Bound}
 A unified expression of the minimax rates for robust estimation was developed by~\citet{chen2018robustcov}. The minimax rate is defined as the quantity $M(\epsilon)$ that satisfies, for a constant $c>0$,
 \begin{equation}\label{eq:rsdefminmaxrate}
 \inf_{\hat{\theta}}\sup_{\overline{\PP}\in\mathcal{A}_\epsilon(\PP_\star)}\overline{\PP}\left(\EE_{\PP_\star}[\ell(\hat\theta(\hat \PP_n), \xi)]\geq M(\epsilon)\right)\geq c
 \end{equation}
 holds. Here the outer probability $\overline{\PP}$ (shorthand for the product measure $\overline{\PP}^n$) is over the randomness of the draw of the contaminated samples $\hat \PP_n$ and the inner expectation $\EE_{\PP_\star}$ denotes averaging over the draw of $\xi$ following the true data-generating distribution $\PP_\star$. This minimax rate was shown to have the form of
 \[ M(\epsilon) = \max\{M(0), \omega(\epsilon, \mathcal{F})\},\]
 where $M(0)$ is the classical minimax rate for uncontaminated distributions, and $\omega(\epsilon,\mathcal{F})$ represents the modulus of continuity over a family $\mathcal{F}$ of probability distributions. For example, $\mc F$ represents the convex combination of normal distributions and an $\epsilon$ fraction of contamination in Huber's contamination model.

This concept of modulus of continuity, tracing back to the foundational work of Donoho and Liu~\citep{donoho1991geometrizing,donoho1994statistical}, represents the fact that in the worst-case contamination scenario, it is theoretically impossible to distinguish between parameters within 
$\omega(\epsilon, \mathcal{F})$ for a given loss.


In the existence of densities,~\citet{liu2019density} studied density estimation under pointwise loss in the presence of contamination, and derived the minimax optimal rate.

Besides the task of robust mean estimation, recent works have also focused on robust covariance estimation~\citep{lai2016agnostic,diakonikolas2019robust,zhu2022generalized},  learning mixtures of spherical Gaussians~\citep{kothari2017better,hopkins2018mixture}, lower bounds against statistical query algorithms~\citep{diakonikolas}, list-decodable learning~\citep{diakonikolas2018list,karmalkar2019list,raghavendra2020list}, robust linear regression~\citep{bhatia2015robust,bhatia2017consistent,klivans2018efficient,suggala2019adaptive} and robust stochastic optimization~\citep{charikar2017learning,diakonikolas2019sever, prasad2018robust}. It is worth noting that an expanding body of research on robust estimation is focusing on robustifying estimators to heavy-tailed distributions, see~\citet{jean2011robustlinear,minsker2015geometric,Donoho2016highdim,devroye2016sub,joly2017on,lugosi2019subgaussian}. The results of which are of a different nature  comparing to the setting of contamination models.

\subsection{Comparing DRO and Robust Statistics}\label{sec:compdroandrobuststat} The attentive reader may recognize that the formulations~\eqref{eq:ERM2} and~\eqref{eq:erm2} presented above is analogous to equations~\eqref{eq:ERM} and~\eqref{eq:erm}. This parallelism is an intentional choice to facilitate a comparison between the frameworks of robust statistics and DRO. We now integrate robust statistics in a lense that is reminiscent of the DRO framework. 

Consider a two-player zero-sum game where the nature's action is to generate the contaminated in-sample data set $\hat \PP_n$, and, having observed $\hat \PP_n$, the statistician's action space is a class of policies $\Psi$ for constructing an estimator $\hat\theta(\cdot): \mc P(\Xi) \rightarrow \Theta$. The loss incurred to the statistician (or equivalently the gain incurred to the nature) is thus
\[
\EE_{\widetilde \PP}[\ell(\hat\theta(\hat \PP_n), \xi)],
\]
which measures the risk of the chosen decision $\hat\theta(\hat \PP_n)$ in the out-of-sample environment $\widetilde \PP$. In the context of robust statistics, this environment is the same as the original data-generating distribution, i.e.~$\widetilde \PP = \PP_\star$. The important characterization of this game is that the nature first generates $\hat \PP_n$, after which the statistician chooses the decision which takes advantage of the realizations $\hat \PP_n$. Hence, we have a \textit{max-min} game 
\begin{equation}\label{eq:robuststatgame}
\sup_{\overline{\PP}\in\mathcal{A}_\epsilon(\PP_\star)}\inf_{\hat\theta(\cdot)\in\Psi}\EE_{\overline{\PP}}\left[\EE_{\PP_\star}[\ell(\hat\theta(\hat \PP_n), \xi)]\right],
\end{equation}
where the outer expectation $\EE_{\overline{\PP}}$ averages the randomness over the draw of the contaminated samples $\hat \PP_n$ and the inner expectation $\EE_{\PP_\star}$  denotes averaging over the draw of $\xi$ following the true data-generating distribution $\PP_\star$.

This is a harder game to play for the nature given the ordering of the actions, as the nature needs to foresee the choice of the statistician. When $\hat \PP_n$ is generated according to the adaptive contamination model, the game~\eqref{eq:robuststatgame} is framed as the sample-wise max-min game
\begin{equation}\label{eq:adarobuststatgame}
\sup_{\hat \PP_n\in\mathcal{A}_\epsilon({\PP}_n)}\inf_{\hat\theta(\cdot)\in\Psi}\EE_{\PP_\star}[\ell(\hat\theta(\hat \PP_n), \xi)],
\end{equation}
whose value is dependent on the realization ${\PP}_n$ of the empirical distribution of uncontaminated samples. 

Comparing this to the DRO formulation~\eqref{eq:DRO}, which is a \textit{min-max} game repeated below with a slightly different presentation :
\begin{equation}\label{eq:DROgame}
    \inf_{\hat \theta(\cdot)\in \Phi} \sup_{\widetilde{\PP} \in \mc B_{\delta}(\hat \PP_n)} \EE_{\widetilde\PP}[\ell(\hat \theta(\hat \PP_n), \xi)].
\end{equation}
In this game, the statistician first observes the clean in-sample data set $\hat\PP_n$, and makes a decision $\hat \theta(\hat\PP_n)$ according to a class of policies $\Phi$. After that, nature enters into the game by choosing an out-of-sample environment $\widetilde \PP \in \mc B_{\delta}(\hat \PP_n) = \{\PP: \mathds{D}(\PP, \hat \PP_n) \leq \delta\}$ with some probability metric $\mathds{D}$, in an adversarial fashion relative to the statistician's decision. In particular, nature's action can depart from the in-sample empirical measure $\hat\PP_n$, and is constrained by the ambiguity set $\mc B(\hat \PP_n)$. The loss incurred to the statistician (or equivalently the gain incurred to the nature) is thus
\[
\EE_{\widetilde{\PP}}[\ell(\hat \theta(\hat\PP_n), \xi)],
\]
which measures the risk of the decision $\hat \theta(\hat\PP_n)$ in the out-of-sample environment $\widetilde{\PP}$. This is a harder game to play for the statistician given the ordering of the actions. 

The value of the game~\eqref{eq:DROgame} is also sample dependent, which is comparable to the game~\eqref{eq:adarobuststatgame} under the adaptive contamination model. A slight adjustment to this DRO formulation would bear a closer resemblance with~\eqref{eq:robuststatgame}. Here we consider the formulation

\[
\inf_{\hat \theta(\cdot)\in\Phi} \sup_{\widetilde \PP \in \mc B_{\delta}(\PP_\star)} \EE_{\PP_\star}\left[\EE_{\widetilde \PP}[\ell(\hat \theta(\hat \PP_n), \xi)]\right],
\]
where $\mc B_{\delta}(\PP_{\star})$ is the set of adversarial distributions. The outer expectation $\EE_{\PP_\star}$ is averaging over the randomness of the empirical measure $\hat\PP_n$, upon which the statistician's action $\hat \theta(\cdot)\in\Phi$ is based, and the inner expectation $\EE_{\widetilde{\PP}}$ averages the out-of-sample environment chosen by the nature.

\subsection{Connection of Robust Statistics to Rockafellian relaxations}
In the context of robust statistics, once the statistician receives the contaminated sample $\hat \PP_n$, a sensible strategy is first to rectify the contamination, possibly in an ``optimistic'' fashion, and then optimize the loss function based on the rectified data points. 
This decision-making approach involves a joint minimization of the form
\begin{equation}\label{eq:minmin}
\inf_{\theta\in\Theta}\inf_{\QQ\in \mc R(\hat \PP_n)} \EE_{\QQ}[\ell(\theta, \xi)].
\end{equation}
Recent studies, such as those by 
\citet{norton2017optimistic,nguyen2019optimistic,jiang2021dfo,royset2023rockafellian,gotoh2023datadriven}, have considered this optimistic formulation in problem settings where inherent conservativeness of the DRO paradigm could render an actually optimal solution as suboptimal~\citep{royset2023rockafellian}. For example, \citet{nguyen2019optimistic} considers an optimistic formulation where the goal is to find the best-case measure that maximizes the non-parametric likelihood in the setting of Bayesian likelihood approximation, and the rectification set $\mc R$ is constrained by KL-divergences, moment constraints, or Wasserstein distances. Also see \citet{gotoh2023datadriven} for a result that the optimistic decision-making can sometimes outperform the empirical risk minimizer in out-of-sample contexts while  there is no guarantee for DRO solutions. Comparing to DRO, the optimistic formulation is typically non-convex,  see \citet{aravkin2020trimmed} for a stochastic proximal-gradient algorithm for solving the resulting nonconvex optimization problem in the setting of trimmed $M$-estimators.

The concept of optimistic decision-making in the operations research literature can be traced back to ``Rockafellian Relaxations''~\citep{rockafellianthesis,rockafellar1974conjugate,ROCKAFELLAR1985extension, rockafellar1997convex,royset2021good,royset2022optimization,royset2023rockafellian}. Also refer to~\citet{zalinescu2002convex,bauschke2011convex}, where optimistic decision making was studied under the name ``perturbation functions''
or ``bivariate functions''.

A natural question is how to choose the rectification set $\mc R (\hat \PP_n)$ appropriately to clean up the contamination. 
In recent works, it is shown that there are some interesting connections between~\eqref{eq:minmin} and well-known estimator in robust statistics. For example, in the setting of univariate mean estimation,~\citet{jiang2021dfo} considers the simple formulation
\[
\inf_{\theta}\inf_{\boldsymbol{\omega}\in\mc R} \sum_{i=1}^n \omega_i(\theta-\xi_i)^2,
\]
where the rectification set is defined as a re-weighting of the samples
\[
\mc R = \left\{\boldsymbol{\omega}\in\R^n_+: \sum_{i=1}^n\frac{1}{\omega_i} = n^2\right\}.
\]
They show that the resulting estimator corresponds to the sample median. This rectification set is slightly more general than formulation~\eqref{eq:minmin} in that the weights $\omega_i$ do not necessarily sum to one. Nevertheless, under this rectification set, the authors~\citet{jiang2021dfo} also show that they are able to recover more similar robust statistics, such
as median absolute deviation, least absolute deviation, and least median of squares.


We conclude this section by demonstrating that the solution to the min-min formulation~\eqref{eq:minmin} recovers an instance of statistical minimax optimality in the literature.
\begin{theorem}[{Informal corollary of \citep[Theorem H.1]{zhu2022generalized}}]\label{thm:minminminimax}
Let $\PP_\star$ be $\sigma^2$-subgaussian, let $\overline{\PP}\in\mc A_\epsilon(\PP_\star)$ follow the $\epsilon$-total variation contamination model, from which the contaminated sample $\hat \PP_n$ are drawn~i.i.d, and let $\mc R_\delta(\hat\PP_n)$ be a $\delta$-total variation neighborhood of $\hat \PP_n$ comprising ``resilient'' distributions \citep[equation (441)]{zhu2022generalized}. 
        Then, choosing $\delta  =2\left(\sqrt{\epsilon} + \sqrt{\frac{\log(1/\eta)}{2n}}\right)^2$, we define the estimator $\hat\theta(\hat {\mathbb{P}}_n)$ as the solution to
        \[
        \mathop{\arg\min}_{\theta\in\R^d} \min_{\mathbb{Q}\in\mc R_\delta(\hat {\mathbb{P}}_n)}\mathbb{E}_{\mathbb{Q}}\left[\left\|\theta-\xi\right\|^2_2\right].
        \]
        With  probability at least $1-3\eta$, it holds that
        \begin{align*}
            \left\|\theta(\mathbb{P}_\star) - \hat\theta(\hat {\mathbb{P}}_n)\right\|_2 
            \leq C\sigma\left(\epsilon\sqrt{\log(\epsilon)} + \sqrt{\frac{d+\log(1/\eta)}{n}}\right),
        \end{align*}
        where $C$ is a universal constant. The dependence on $\epsilon,d$ and $n$ are information-theoretically optimal (cf. equation~\eqref{eq:rsdefminmaxrate}).
\end{theorem}

We remark that minimax optimality of the min-min formulation~\eqref{eq:minmin} in a wider context (e.g., Wasserstein contamination) is a prospective area of interest for future research. 


\section{Conclusions and Discussion}

The goal of this section is to briefly discuss various areas of potential research interest in connection with DRO and statistics. The discussion is not exhaustive, but rather we want to expose the fact that DRO as a statistical tool offers a wide range of opportunities for the statistical community.

We mentioned in Section~\ref{Subsubsec:asymnormal}, simply in terms of asymptotic mean squared error as the sample size increases, it is sensible to ponder on the benefits of DRO estimators. The work of~\citet{lam2021impossibility} shows that in the presence of sufficient regularity (e.g. smoothness of the loss) DRO estimators tend to dominate in second-order stochastic dominance of the empirical risk minimization estimator of the optimal loss. This observation is consistent with our discussion in Section \ref{Subsubsec:asymnormal}. Nevertheless, the situation may be different if these regularity conditions are not satisfied. For instance, \citet[Section 3.3]{duchi2018variance} shows that in settings involving non-smooth losses, DRO estimators may enjoy superior rates compared to their empirical risk minimization counterparts. 

Continuing in the context of classical statistical analysis involving large sample properties. There are objects that the DRO estimation approach offers that are interesting statistically speaking. The most natural such object is the worst-case distribution, which is a by-product of the DRO approach and possesses rich interpretations. Even in the context of the estimators that DRO recovers exactly and that are well-known in statistics, the DRO approach furnishes additional insight to these classical estimators using the associated worst-case distribution. 

Another example of an interesting statistical object to study offered by DRO formulations is the natural confidence region induced by the DRO and discussed in Subsection \ref{subsect:confidence}. Using the duality between confidence regions and hypothesis testing we can compare the efficiency of various confidence regions implied by standard notions of efficiency in hypothesis testing. 

Statistical efficiency is also of interest in connection to important parameters, such as the dimension, for example. We have seen that a suitably chosen distributional uncertainty region can be used to show the equivalence between a DRO estimator and a well-known estimator. An example of this situation is square-root LASSO and the DRO-motivated choice of uncertainty size recovers regularization prescriptions studied in the high-dimensional statistics literature. Likewise, in the context of $\phi$-divergence, the DRO-based estimator is used to re-weight samples in order to hedge against significant inference errors in subpopulations. In summary, while the DRO estimator may have a higher asymptotic mean squared error compared to the empirical risk minimization estimator when used in situations in which the statistical problem is ill-posed (i.e. the sample size is relatively small compared to the information required to estimate the parameter) or when the goal is not purely based on mean-squared error but we are interested in hedging a different type of risk, then DRO based estimation offers enough flexibility and interpretability, not only through their formulation but also through the associated worst-case distribution. 

In general, we also note that adding constraints or exploring other types of distributional uncertainty sets that can be used to better inform the attributes of the adversary to reduce conservativeness is a significant topic of research interest. For example, the work of~\citet{olea2022generalization} explores different DRO uncertainty sets based on the sliced Wasserstein distance. The advantage of this formulation is that it does not suffer from the statistical course of dimensionality for comparing distributions in high dimensions (as is the case of the Wasserstein distance); see also the approaches recently advocated by \citet{bennouna2022holistic, liu2023smoothed}.

Another area of significant interest which we touched only superficially is the issue of fairness. We mentioned that $\phi$-divergence has been utilized to try to improve the inference quality in estimated statistics involving minority sub-populations. Other DRO-based ideas have been recently applied in the context of fairness. For example, \citet{taskesen2020distributionally, si2021testing} propose a projection-based hypothesis test closely related to the one discussed in Section \ref{subsect:confidence} for algorithmic fairness. This is a setting in which the associated distribution induced by DRO-type mechanisms deserves significantly more statistical investigation.

Next, we comment on dynamic DRO settings. This is an area that closely connects with what is known as distributionally robust reinforcement learning and it is in its infancy (see, e.g., \citet{xu2010distributionally, osogami2012robustness, lim2013reinforcement,zhou2021finite,backhoff2022estimating,si2023distributionally}). Even fundamental problems involving how to formulate associated distributionally robust Markov decision processes based on the sequentially available information for the agent and the adversary are significantly non-trivial (see~\citet{wang2023foundation}). This area opens up a wide range of interesting questions for the statistics community. To give a sense of why DRO naturally offers a meaningful approach to estimation and optimization in these settings, note that in many situations of interest in stochastic control, there is a real possibility of facing unobserved (i.e. confounding) variables. This type of formulation is naturally posed as a so-called Partially Observed Markov Decision Process, which is challenging to study since it requires a history-dependent specification at every point in time. In these settings, the statistician can introduce a Markovian model (thus reducing the problem to a standard reinforcement learning environment) and instead use DRO to hedge against the model misspecification which has been introduced for tractability. 

Finally, we finish our discussion by noting that the robust statistics perspective offered in this paper provides a useful point of view to connect and contrast DRO estimators and classical robust estimators. This perspective, characterized by the order in which the statistician and the adversary make their decision, was introduced in this paper primarily to motivate the fundamental differences in the nature of these types of robust estimators, The DRO estimator is pessimistic in nature because the statistician is at the mercy of an adversary that will change the out-of-sample environment. In robust statistics, hidden in the data lies useful information about the actual out-of-sample distribution - the adversary has made its move. Therefore, the statistician naturally could try to clean or rectify the contamination employed by the adversary thus leading to an optimistic approach. 

\section*{Acknowledgments} 
The material in this paper is based upon work supported by the Air Force Office of Scientific Research under award number FA9550-20-1-0397. Additional support is gratefully acknowledged from NSF 1915967, 2118199, 2229012, 2312204.

\newpage 
\bibliography{sts_submission/bibliography}

\begin{thebibliography}{}

\bibitem[An and Gao, 2021]{an2021generalization}
An, Y. and Gao, R. (2021).
\newblock Generalization bounds for ({W}asserstein) robust optimization.
\newblock In {\em Advances in Neural Information Processing Systems}, volume~34, pages 10382--10392.

\bibitem[Aravkin and Davis, 2020]{aravkin2020trimmed}
Aravkin, A. and Davis, D. (2020).
\newblock Trimmed statistical estimation via variance reduction.
\newblock {\em Mathematics of Operations Research}, 45(1):292--322.

\bibitem[Audibert and Catoni, 2011]{jean2011robustlinear}
Audibert, J.-Y. and Catoni, O. (2011).
\newblock {Robust linear least squares regression}.
\newblock {\em Annals of Statistics}, 39(5):2766 -- 2794.

\bibitem[Azizian et~al., 2023]{azizian2023regularization}
Azizian, W., Iutzeler, F., and Malick, J. (2023).
\newblock Regularization for wasserstein distributionally robust optimization.
\newblock {\em ESAIM: Control, Optimisation and Calculus of Variations}, 29:33.

\bibitem[Backhoff et~al., 2022]{backhoff2022estimating}
Backhoff, J., Bartl, D., Beiglb{\"o}ck, M., and Wiesel, J. (2022).
\newblock Estimating processes in adapted {W}asserstein distance.
\newblock {\em Annals of Applied Probability}, 32(1):529--550.

\bibitem[Bartl et~al., 2021]{bartl2021sensitivity}
Bartl, D., Drapeau, S., Ob{\l}{\'o}j, J., and Wiesel, J. (2021).
\newblock Sensitivity analysis of {W}asserstein distributionally robust optimization problems.
\newblock {\em Proceedings of the Royal Society A}, 477(2256):20210176.

\bibitem[Bateni and Dalalyan, 2019]{Bateni2019ConfidenceRA}
Bateni, A.-H. and Dalalyan, A.~S. (2019).
\newblock Confidence regions and minimax rates in outlier-robust estimation on the probability simplex.
\newblock {\em Electronic Journal of Statistics}.

\bibitem[Bauschke and Combettes, 2011]{bauschke2011convex}
Bauschke, H. and Combettes, P. (2011).
\newblock {\em Convex Analysis and Monotone Operator Theory in Hilbert Spaces}.
\newblock CMS Books in Mathematics. Springer New York.

\bibitem[Belloni et~al., 2011]{belloni2011square}
Belloni, A., Chernozhukov, V., and Wang, L. (2011).
\newblock Square-root lasso: pivotal recovery of sparse signals via conic programming.
\newblock {\em Biometrika}, 98(4):791--806.

\bibitem[Bennett et~al., 2023]{bennett2023minimax}
Bennett, A., Kallus, N., Mao, X., Newey, W., Syrgkanis, V., and Uehara, M. (2023).
\newblock Minimax instrumental variable regression and {$L_2$} convergence guarantees without identification or closedness.
\newblock {\em arXiv preprint arXiv:2302.05404}.

\bibitem[Bennouna and Van~Parys, 2022]{bennouna2022holistic}
Bennouna, A. and Van~Parys, B. (2022).
\newblock Holistic robust data-driven decisions.
\newblock {\em arXiv preprint arXiv:2207.09560}.

\bibitem[Berlinet and Thomas-Agnan, 2011]{berlinet2011reproducing}
Berlinet, A. and Thomas-Agnan, C. (2011).
\newblock {\em Reproducing Kernel Hilbert Spaces in Probability and Statistics}.
\newblock Springer Science \& Business Media.

\bibitem[Bernholt, 2006]{bernholt2006robust}
Bernholt, T. (2006).
\newblock Robust estimators are hard to compute.
\newblock Technical Report 2005,52, Universität Dortmund.

\bibitem[Bertsimas et~al., 2022]{bertsimas2022distributionally}
Bertsimas, D., Imai, K., and Li, M.~L. (2022).
\newblock Distributionally robust causal inference with observational data.
\newblock {\em arXiv preprint arXiv:2210.08326}.

\bibitem[Bhatia et~al., 2017]{bhatia2017consistent}
Bhatia, K., Jain, P., Kamalaruban, P., and Kar, P. (2017).
\newblock Consistent robust regression.
\newblock In {\em Advances in Neural Information Processing Systems}, volume~30.

\bibitem[Bhatia et~al., 2015]{bhatia2015robust}
Bhatia, K., Jain, P., and Kar, P. (2015).
\newblock Robust regression via hard thresholding.
\newblock In {\em Advances in Neural Information Processing Systems}, volume~28.

\bibitem[Blanchet and Kang, 2017]{blanchet2017distributionally}
Blanchet, J. and Kang, Y. (2017).
\newblock Distributionally robust groupwise regularization estimator.
\newblock In {\em Asian Conference on Machine Learning}, pages 97--112. PMLR.

\bibitem[Blanchet and Kang, 2020]{blanchet2020semi}
Blanchet, J. and Kang, Y. (2020).
\newblock Semi-supervised learning based on distributionally robust optimization.
\newblock {\em Data Analysis and Applications 3: Computational, Classification, Financial, Statistical and Stochastic Methods}, 5:1--33.

\bibitem[Blanchet and Kang, 2021]{blanchet2021sample}
Blanchet, J. and Kang, Y. (2021).
\newblock Sample out-of-sample inference based on {W}asserstein distance.
\newblock {\em Operations Research}, 69(3):985--1013.

\bibitem[Blanchet et~al., 2019a]{blanchet2019robust}
Blanchet, J., Kang, Y., and Murthy, K. (2019a).
\newblock Robust {W}asserstein profile inference and applications to machine learning.
\newblock {\em Journal of Applied Probability}, 56(3):830--857.

\bibitem[Blanchet et~al., 2023a]{blanchet2020machine}
Blanchet, J., Kang, Y., Olea, J. L.~M., Nguyen, V.~A., and Zhang, X. (2023a).
\newblock Dropout training is distributionally robust optimal.
\newblock {\em Journal of Machine Learning Research}, 24(180):1--60.

\bibitem[Blanchet et~al., 2021a]{blanchet2021doubly}
Blanchet, J., Kang, Y., Zhang, F., He, F., and Hu, Z. (2021a).
\newblock Doubly robust data-driven distributionally robust optimization.
\newblock {\em Applied Modeling Techniques and Data Analysis 1: Computational Data Analysis Methods and Tools}, 7:75--90.

\bibitem[Blanchet et~al., 2023b]{blanchet2023unifying}
Blanchet, J., Kuhn, D., Li, J., and Taskesen, B. (2023b).
\newblock Unifying distributionally robust optimization via optimal transport theory.
\newblock {\em arXiv preprint arXiv:2308.05414}.

\bibitem[Blanchet et~al., 2021b]{blanchet2021statistical}
Blanchet, J., Murthy, K., and Nguyen, V.~A. (2021b).
\newblock Statistical analysis of wasserstein distributionally robust estimators.
\newblock In {\em Tutorials in Operations Research: Emerging Optimization Methods and Modeling Techniques with Applications}, pages 227--254. INFORMS.

\bibitem[Blanchet et~al., 2022a]{blanchet2022confidence}
Blanchet, J., Murthy, K., and Si, N. (2022a).
\newblock Confidence regions in {W}asserstein distributionally robust estimation.
\newblock {\em Biometrika}, 109(2):295--315.

\bibitem[Blanchet et~al., 2022b]{blanchet2022optimal}
Blanchet, J., Murthy, K., and Zhang, F. (2022b).
\newblock Optimal transport-based distributionally robust optimization: Structural properties and iterative schemes.
\newblock {\em Mathematics of Operations Research}, 47(2):1500--1529.

\bibitem[Blanchet and Shapiro, 2023]{blanchet2023statistical}
Blanchet, J. and Shapiro, A. (2023).
\newblock Statistical limit theorems in distributionally robust optimization.
\newblock {\em arXiv preprint arXiv:2303.14867}.

\bibitem[Blanchet et~al., 2019b]{blanchet2019distributionally}
Blanchet, J., Zhang, F., Kang, Y., and Hu, Z. (2019b).
\newblock A distributionally robust boosting algorithm.
\newblock In {\em 2019 Winter Simulation Conference}, pages 3728--3739. IEEE.

\bibitem[Box, 1976]{box1976science}
Box, G.~E. (1976).
\newblock Science and statistics.
\newblock {\em Journal of the American Statistical Association}, 71(356):791--799.

\bibitem[Box, 1979]{BOX1979robustness}
Box, G.~E. (1979).
\newblock Robustness in the strategy of scientific model building.
\newblock In {\em Robustness in Statistics}, pages 201--236. Elsevier.

\bibitem[Box, 1953]{box1953nonnormality}
Box, G. E.~P. (1953).
\newblock Non-normality and tests on variances.
\newblock {\em Biometrika}, 40(3/4):318--335.

\bibitem[Chao et~al., 2019]{chao2018robust}
Chao, G., Yuan, Y., and Weizhi, Z. (2019).
\newblock Robust estimation via generative adversarial networks.
\newblock In {\em International Conference on Learning Representations}.

\bibitem[Charikar et~al., 2017]{charikar2017learning}
Charikar, M., Steinhardt, J., and Valiant, G. (2017).
\newblock Learning from untrusted data.
\newblock In {\em Proceedings of the 49th Annual ACM SIGACT Symposium on Theory of Computing}, pages 47--60.

\bibitem[Chen et~al., 2018]{chen2018robustcov}
Chen, M., Gao, C., and Ren, Z. (2018).
\newblock {Robust covariance and scatter matrix estimation under Huber’s contamination model}.
\newblock {\em Annals of Statistics}, 46(5):1932 -- 1960.

\bibitem[Csisz{\'a}r, 1975]{csiszar1975divergence}
Csisz{\'a}r, I. (1975).
\newblock I-divergence geometry of probability distributions and minimization problems.
\newblock {\em Annals of Probability}, pages 146--158.

\bibitem[Dapogny et~al., 2023]{dapogny2023entropy}
Dapogny, C., Iutzeler, F., Meda, A., and Thibert, B. (2023).
\newblock Entropy-regularized wasserstein distributionally robust shape and topology optimization.
\newblock {\em Structural and Multidisciplinary Optimization}, 66(3):42.

\bibitem[Delage and Ye, 2010]{delage2010distributionally}
Delage, E. and Ye, Y. (2010).
\newblock Distributionally robust optimization under moment uncertainty with application to data-driven problems.
\newblock {\em Operations Research}, 58(3):595--612.

\bibitem[Depersin and Lecu{\'e}, 2022]{depersin2022robust}
Depersin, J. and Lecu{\'e}, G. (2022).
\newblock {Robust sub-Gaussian estimation of a mean vector in nearly linear time}.
\newblock {\em Annals of Statistics}, 50(1):511 -- 536.

\bibitem[Devroye et~al., 2016]{devroye2016sub}
Devroye, L., Lerasle, M., Lugosi, G., and Oliveira, R.~I. (2016).
\newblock {Sub-Gaussian mean estimators}.
\newblock {\em Annals of Statistics}, 44(6):2695 -- 2725.

\bibitem[Diakonikolas et~al., 2019a]{diakonikolas2019robust}
Diakonikolas, I., Kamath, G., Kane, D., Li, J., Moitra, A., and Stewart, A. (2019a).
\newblock Robust estimators in high-dimensions without the computational intractability.
\newblock {\em SIAM Journal on Computing}, 48(2):742--864.

\bibitem[Diakonikolas et~al., 2019b]{diakonikolas2019sever}
Diakonikolas, I., Kamath, G., Kane, D., Li, J., Steinhardt, J., and Stewart, A. (2019b).
\newblock Sever: A robust meta-algorithm for stochastic optimization.
\newblock In {\em International Conference on Machine Learning}, pages 1596--1606. PMLR.

\bibitem[Diakonikolas et~al., 2017a]{diakonikolas2017being}
Diakonikolas, I., Kamath, G., Kane, D.~M., Li, J., Moitra, A., and Stewart, A. (2017a).
\newblock Being robust (in high dimensions) can be practical.
\newblock In {\em International Conference on Machine Learning}, pages 999--1008. PMLR.

\bibitem[Diakonikolas and Kane, 2023]{diakonikolas2023algorithmic}
Diakonikolas, I. and Kane, D. (2023).
\newblock {\em Algorithmic High-Dimensional Robust Statistics}.
\newblock Cambridge University Press.

\bibitem[Diakonikolas and Kane, 2019]{diakonikolas2019recent}
Diakonikolas, I. and Kane, D.~M. (2019).
\newblock Recent advances in algorithmic high-dimensional robust statistics.
\newblock {\em arXiv preprint arXiv:1911.05911}.

\bibitem[Diakonikolas et~al., 2017b]{diakonikolas}
Diakonikolas, I., Kane, D.~M., and Stewart, A. (2017b).
\newblock Statistical query lower bounds for robust estimation of high-dimensional {G}aussians and {G}aussian mixtures.
\newblock In {\em 2017 IEEE 58th Annual Symposium on Foundations of Computer Science}, pages 73--84.

\bibitem[Diakonikolas et~al., 2018a]{diakonikolas2018learning}
Diakonikolas, I., Kane, D.~M., and Stewart, A. (2018a).
\newblock Learning geometric concepts with nasty noise.
\newblock In {\em Proceedings of the 50th Annual ACM SIGACT Symposium on Theory of Computing}, pages 1061--1073.

\bibitem[Diakonikolas et~al., 2018b]{diakonikolas2018list}
Diakonikolas, I., Kane, D.~M., and Stewart, A. (2018b).
\newblock List-decodable robust mean estimation and learning mixtures of spherical gaussians.
\newblock In {\em Proceedings of the 50th Annual ACM SIGACT Symposium on Theory of Computing}, pages 1047--1060.

\bibitem[Donoho and Montanari, 2016]{Donoho2016highdim}
Donoho, D. and Montanari, A. (2016).
\newblock High dimensional robust {M}-estimation: {A}symptotic variance via approximate message passing.
\newblock {\em Probability Theory and Related Fields}, 166(3):935--969.

\bibitem[Donoho, 1994]{donoho1994statistical}
Donoho, D.~L. (1994).
\newblock {Statistical Estimation and Optimal Recovery}.
\newblock {\em Annals of Statistics}, 22(1):238 -- 270.

\bibitem[Donoho and Gasko, 1992]{donoho1992breakdown}
Donoho, D.~L. and Gasko, M. (1992).
\newblock Breakdown properties of location estimates based on halfspace depth and projected outlyingness.
\newblock {\em Annals of Statistics}, 20(4):1803--1827.

\bibitem[Donoho and Huber, 1983]{donoho1983notion}
Donoho, D.~L. and Huber, P.~J. (1983).
\newblock The notion of breakdown point.
\newblock {\em A festschrift for Erich L. Lehmann}, 157184.

\bibitem[Donoho and Liu, 1988]{donoho1988automatic}
Donoho, D.~L. and Liu, R.~C. (1988).
\newblock The ``automatic" robustness of minimum distance functionals.
\newblock {\em Annals of Statistics}, 16(2):552--586.

\bibitem[Donoho and Liu, 1991]{donoho1991geometrizing}
Donoho, D.~L. and Liu, R.~C. (1991).
\newblock Geometrizing rates of convergence, {III}.
\newblock {\em Annals of Statistics}, pages 668--701.

\bibitem[Duchi et~al., 2023]{duchi2023distributionally}
Duchi, J., Hashimoto, T., and Namkoong, H. (2023).
\newblock Distributionally robust losses for latent covariate mixtures.
\newblock {\em Operations Research}, 71(2):649--664.

\bibitem[Duchi and Namkoong, 2018]{duchi2018variance}
Duchi, J. and Namkoong, H. (2018).
\newblock Variance-based regularization with convex objectives.
\newblock {\em Journal of Machine Learning Research}, 19:1--55.

\bibitem[Duchi et~al., 2021]{duchi2021statistics}
Duchi, J.~C., Glynn, P.~W., and Namkoong, H. (2021).
\newblock Statistics of robust optimization: A generalized empirical likelihood approach.
\newblock {\em Mathematics of Operations Research}, 46(3):946--969.

\bibitem[Duchi and Namkoong, 2021]{duchi2021learning}
Duchi, J.~C. and Namkoong, H. (2021).
\newblock Learning models with uniform performance via distributionally robust optimization.
\newblock {\em Annals of Statistics}, 49(3):1378--1406.

\bibitem[Fournier and Guillin, 2015]{fournier2015rate}
Fournier, N. and Guillin, A. (2015).
\newblock On the rate of convergence in {W}asserstein distance of the empirical measure.
\newblock {\em Probability Theory and Related Fields}, 162(3-4):707--738.

\bibitem[Freund and Schapire, 1997]{freund1997decision}
Freund, Y. and Schapire, R.~E. (1997).
\newblock A decision-theoretic generalization of on-line learning and an application to boosting.
\newblock {\em Journal of Computer and System Sciences}, 55(1):119--139.

\bibitem[Gao, 2020]{gao2020robustregression}
Gao, C. (2020).
\newblock {Robust regression via mutivariate regression depth}.
\newblock {\em Bernoulli}, 26(2):1139 -- 1170.

\bibitem[Gao, 2022]{gao2022finite}
Gao, R. (2022).
\newblock Finite-sample guarantees for {W}asserstein distributionally robust optimization: Breaking the curse of dimensionality.
\newblock {\em Operations Research}.

\bibitem[Gao et~al., 2022]{gao2022wasserstein}
Gao, R., Chen, X., and Kleywegt, A.~J. (2022).
\newblock {W}asserstein distributionally robust optimization and variation regularization.
\newblock {\em Operations Research}.

\bibitem[Gao and Kleywegt, 2023]{gao2023distributionally}
Gao, R. and Kleywegt, A. (2023).
\newblock Distributionally robust stochastic optimization with {W}asserstein distance.
\newblock {\em Mathematics of Operations Research}, 48(2):603--655.

\bibitem[Gao et~al., 2018]{gao2018robust}
Gao, R., Xie, L., Xie, Y., and Xu, H. (2018).
\newblock Robust hypothesis testing using {W}asserstein uncertainty sets.
\newblock In {\em Advances in Neural Information Processing Systems}, volume~31.

\bibitem[Goodfellow et~al., 2014]{goodfellow2014generative}
Goodfellow, I., Pouget-Abadie, J., Mirza, M., Xu, B., Warde-Farley, D., Ozair, S., Courville, A., and Bengio, Y. (2014).
\newblock Generative adversarial nets.
\newblock In {\em Advances in Neural Information Processing Systems}, volume~27.

\bibitem[Goodfellow et~al., 2015]{goodfellow2014explaining}
Goodfellow, I.~J., Shlens, J., and Szegedy, C. (2015).
\newblock Explaining and harnessing adversarial examples.
\newblock In {\em International Conference on Learning Representations}.

\bibitem[Gotoh et~al., 2023]{gotoh2023datadriven}
Gotoh, J.-y., Kim, M.~J., and Lim, A.~E. (2023).
\newblock A data-driven approach to beating {SAA} out of sample.
\newblock {\em Operations Research}.

\bibitem[G{\"u}l and Zoubir, 2017]{gul2017minimax}
G{\"u}l, G. and Zoubir, A.~M. (2017).
\newblock Minimax robust hypothesis testing.
\newblock {\em IEEE Transactions on Information Theory}, 63(9):5572--5587.

\bibitem[Hampel, 1968]{hampel1968contributions}
Hampel, F. (1968).
\newblock {\em Contributions to the Theory of Robust Estimation}.
\newblock University of California.

\bibitem[Hampel, 1971]{hampel1971general}
Hampel, F.~R. (1971).
\newblock A general qualitative definition of robustness.
\newblock {\em Annals of Mathematical Statistics}, 42(6):1887 -- 1896.

\bibitem[He and Lam, 2021]{he2021higher}
He, S. and Lam, H. (2021).
\newblock Higher-order expansion and bartlett correctability of distributionally robust optimization.
\newblock {\em arXiv preprint arXiv:2108.05908}.

\bibitem[Hopkins and Li, 2018]{hopkins2018mixture}
Hopkins, S.~B. and Li, J. (2018).
\newblock Mixture models, robustness, and sum of squares proofs.
\newblock In {\em Proceedings of the 50th Annual ACM SIGACT Symposium on Theory of Computing}, pages 1021--1034.

\bibitem[Hu et~al., 2018]{hu2018does}
Hu, W., Niu, G., Sato, I., and Sugiyama, M. (2018).
\newblock Does distributionally robust supervised learning give robust classifiers?
\newblock In {\em International Conference on Machine Learning}, pages 2029--2037. PMLR.

\bibitem[Hu and Hong, 2013]{hu2013kullback}
Hu, Z. and Hong, L.~J. (2013).
\newblock Kullback-leibler divergence constrained distributionally robust optimization.
\newblock {\em Available at Optimization Online}, 1(2):9.

\bibitem[Huber, 2004]{huber2004robust}
Huber, P. (2004).
\newblock {\em Robust Statistics}.
\newblock Wiley Series in Probability and Statistics - Applied Probability and Statistics Section Series. Wiley.

\bibitem[Huber, 1964]{huber1964robust}
Huber, P.~J. (1964).
\newblock Robust estimation of a location parameter.
\newblock {\em Annals of Mathematical Statistics}, 35(1):73--101.

\bibitem[Huber, 1968]{Huber1968confidence}
Huber, P.~J. (1968).
\newblock Robust confidence limits.
\newblock {\em Zeitschrift f{\"u}r Wahrscheinlichkeitstheorie und Verwandte Gebiete}, 10(4):269--278.

\bibitem[Huber, 1972]{huber1972review}
Huber, P.~J. (1972).
\newblock The 1972 {W}ald lecture robust statistics: A review.
\newblock {\em Annals of Mathematical Statistics}, 43(4):1041 -- 1067.

\bibitem[Iman et~al., 2023]{iman2023review}
Iman, M., Arabnia, H.~R., and Rasheed, K. (2023).
\newblock A review of deep transfer learning and recent advancements.
\newblock {\em Technologies}, 11(2):40.

\bibitem[Jiang and Xie, 2023]{jiang2021dfo}
Jiang, N. and Xie, W. (2023).
\newblock Distributionally favorable optimization: A framework for data-driven decision-making with endogenous outliers.
\newblock {\em Available at Optimization Online}.

\bibitem[Johnson and Preparata, 1978]{JOHNSON1978densest}
Johnson, D. and Preparata, F. (1978).
\newblock The densest hemisphere problem.
\newblock {\em Theoretical Computer Science}, 6(1):93--107.

\bibitem[Joly et~al., 2017]{joly2017on}
Joly, E., Lugosi, G., and Oliveira, R.~I. (2017).
\newblock {On the estimation of the mean of a random vector}.
\newblock {\em Electronic Journal of Statistics}, 11(1):440 -- 451.

\bibitem[Kantorovich and Rubinshtein, 1958]{kantorovich1958space}
Kantorovich, L.~V. and Rubinshtein, S. (1958).
\newblock On a space of totally additive functions.
\newblock {\em Vestnik of the St. Petersburg University: Mathematics}, 13(7):52--59.

\bibitem[Karmalkar et~al., 2019]{karmalkar2019list}
Karmalkar, S., Klivans, A., and Kothari, P. (2019).
\newblock List-decodable linear regression.
\newblock In {\em Advances in Neural Information Processing Systems}, volume~32.

\bibitem[Klivans et~al., 2018]{klivans2018efficient}
Klivans, A., Kothari, P.~K., and Meka, R. (2018).
\newblock Efficient algorithms for outlier-robust regression.
\newblock In {\em Conference On Learning Theory}, pages 1420--1430. PMLR.

\bibitem[Kothari and Steinhardt, 2017]{kothari2017better}
Kothari, P.~K. and Steinhardt, J. (2017).
\newblock Better agnostic clustering via relaxed tensor norms.
\newblock {\em arXiv preprint arXiv:1711.07465}.

\bibitem[Kuhn et~al., 2019]{kuhn2019wasserstein}
Kuhn, D., Esfahani, P.~M., Nguyen, V.~A., and Shafieezadeh-Abadeh, S. (2019).
\newblock Wasserstein distributionally robust optimization: Theory and applications in machine learning.
\newblock In {\em Operations Research \& Management Science in the Age of Analytics}, pages 130--166. INFORMS.

\bibitem[Lai et~al., 2016]{lai2016agnostic}
Lai, K.~A., Rao, A.~B., and Vempala, S. (2016).
\newblock Agnostic estimation of mean and covariance.
\newblock In {\em 2016 IEEE 57th Annual Symposium on Foundations of Computer Science}, pages 665--674. IEEE Computer Society.

\bibitem[Lam, 2016]{lam2016robust}
Lam, H. (2016).
\newblock Robust sensitivity analysis for stochastic systems.
\newblock {\em Mathematics of Operations Research}, 41(4):1248--1275.

\bibitem[Lam, 2018]{lam2018sensitivity}
Lam, H. (2018).
\newblock Sensitivity to serial dependency of input processes: A robust approach.
\newblock {\em Management Science}, 64(3):1311--1327.

\bibitem[Lam, 2021]{lam2021impossibility}
Lam, H. (2021).
\newblock On the impossibility of statistically improving empirical optimization: A second-order stochastic dominance perspective.
\newblock {\em arXiv preprint arXiv:2105.13419}.

\bibitem[Lam and Zhou, 2017]{lam2017empirical}
Lam, H. and Zhou, E. (2017).
\newblock The empirical likelihood approach to quantifying uncertainty in sample average approximation.
\newblock {\em Operations Research Letters}, 45(4):301--307.

\bibitem[Lee and Raginsky, 2018]{lee2018minimax}
Lee, J. and Raginsky, M. (2018).
\newblock Minimax statistical learning with {W}asserstein distances.
\newblock {\em Advances in Neural Information Processing Systems}, 31.

\bibitem[Levy and Nikoukhah, 2012]{levy2012robust}
Levy, B.~C. and Nikoukhah, R. (2012).
\newblock Robust state space filtering under incremental model perturbations subject to a relative entropy tolerance.
\newblock {\em IEEE Transactions on Automatic Control}, 58(3):682--695.

\bibitem[Levy et~al., 2020]{levy2020large}
Levy, D., Carmon, Y., Duchi, J.~C., and Sidford, A. (2020).
\newblock Large-scale methods for distributionally robust optimization.
\newblock In {\em Advances in Neural Information Processing Systems}, volume~33, pages 8847--8860.

\bibitem[Li et~al., 2020]{li2020fast}
Li, J., Chen, C., and So, A. M.-C. (2020).
\newblock Fast epigraphical projection-based incremental algorithms for {W}asserstein distributionally robust support vector machine.
\newblock In {\em Advances in Neural Information Processing Systems}, volume~33, pages 4029--4039.

\bibitem[Li et~al., 2019]{li2019first}
Li, J., Huang, S., and So, A. M.-C. (2019).
\newblock A first-order algorithmic framework for distributionally robust logistic regression.
\newblock In {\em Advances in Neural Information Processing Systems}, volume~32.

\bibitem[Li et~al., 2022]{li2022tikhonov}
Li, J., Lin, S., Blanchet, J., and Nguyen, V.~A. (2022).
\newblock Tikhonov regularization is optimal transport robust under martingale constraints.
\newblock In {\em Advances in Neural Information Processing Systems}, volume~35, pages 17677--17689.

\bibitem[Lim et~al., 2013]{lim2013reinforcement}
Lim, S.~H., Xu, H., and Mannor, S. (2013).
\newblock Reinforcement learning in robust {M}arkov decision processes.
\newblock In {\em Advances in Neural Information Processing Systems}, volume~26.

\bibitem[Liu and Gao, 2019]{liu2019density}
Liu, H. and Gao, C. (2019).
\newblock {Density estimation with contamination: minimax rates and theory of adaptation}.
\newblock {\em Electronic Journal of Statistics}, 13(2):3613 -- 3653.

\bibitem[Liu et~al., 2020]{liu2020high}
Liu, L., Shen, Y., Li, T., and Caramanis, C. (2020).
\newblock High dimensional robust sparse regression.
\newblock In {\em International Conference on Artificial Intelligence and Statistics}, pages 411--421. PMLR.

\bibitem[Liu and Loh, 2022]{liu2022robust}
Liu, Z. and Loh, P.-L. (2022).
\newblock {Robust W-GAN-based estimation under Wasserstein contamination}.
\newblock {\em Information and Inference: A Journal of the IMA}, 12(1):312--362.

\bibitem[Liu et~al., 2023]{liu2023smoothed}
Liu, Z., Van~Parys, B.~P., and Lam, H. (2023).
\newblock Smoothed $ f $-divergence distributionally robust optimization: Exponential rate efficiency and complexity-free calibration.
\newblock {\em arXiv preprint arXiv:2306.14041}.

\bibitem[Lotidis et~al., 2023]{lotidis2023wasserstein}
Lotidis, K., Bambos, N., Blanchet, J., and Li, J. (2023).
\newblock Wasserstein distributionally robust linear-quadratic estimation under martingale constraints.
\newblock In {\em International Conference on Artificial Intelligence and Statistics}, pages 8629--8644. PMLR.

\bibitem[Lugosi and Mendelson, 2019]{lugosi2019subgaussian}
Lugosi, G. and Mendelson, S. (2019).
\newblock {Sub-Gaussian estimators of the mean of a random vector}.
\newblock {\em Annals of Statistics}, 47(2):783 -- 794.

\bibitem[Madry et~al., 2017]{madry2017towards}
Madry, A., Makelov, A., Schmidt, L., Tsipras, D., and Vladu, A. (2017).
\newblock Towards deep learning models resistant to adversarial attacks.
\newblock {\em arXiv preprint arXiv:1706.06083}.

\bibitem[Minsker, 2015]{minsker2015geometric}
Minsker, S. (2015).
\newblock {Geometric median and robust estimation in Banach spaces}.
\newblock {\em Bernoulli}, 21(4):2308 -- 2335.

\bibitem[Mohajerin~Esfahani and Kuhn, 2018]{mohajerin2018data}
Mohajerin~Esfahani, P. and Kuhn, D. (2018).
\newblock Data-driven distributionally robust optimization using the {W}asserstein metric: {P}erformance guarantees and tractable reformulations.
\newblock {\em Mathematical Programming}, 171(1-2):115--166.

\bibitem[Ng, 2004]{ng2004feature}
Ng, A.~Y. (2004).
\newblock Feature selection, {$L_1$} vs. {$L_2$} regularization, and rotational invariance.
\newblock In {\em Proceedings of the Twenty-First International Conference on Machine Learning}, page~78.

\bibitem[Nguyen et~al., 2023]{nguyen2023bridging}
Nguyen, V.~A., Shafieezadeh-Abadeh, S., Kuhn, D., and Mohajerin~Esfahani, P. (2023).
\newblock Bridging {B}ayesian and minimax mean square error estimation via {W}asserstein distributionally robust optimization.
\newblock {\em Mathematics of Operations Research}, 48(1):1--37.

\bibitem[Nguyen et~al., 2019]{nguyen2019optimistic}
Nguyen, V.~A., Shafieezadeh~Abadeh, S., Yue, M.-C., Kuhn, D., and Wiesemann, W. (2019).
\newblock Optimistic distributionally robust optimization for nonparametric likelihood approximation.
\newblock In {\em Advances in Neural Information Processing Systems}, volume~32.

\bibitem[Nguyen et~al., 2020a]{nguyen2020robust}
Nguyen, V.~A., Si, N., and Blanchet, J. (2020a).
\newblock Robust {B}ayesian classification using an optimistic score ratio.
\newblock In {\em International Conference on Machine Learning}, pages 7327--7337. PMLR.

\bibitem[Nguyen et~al., 2020b]{nguyen2020distributionally}
Nguyen, V.~A., Zhang, F., Blanchet, J., Delage, E., and Ye, Y. (2020b).
\newblock Distributionally robust local non-parametric conditional estimation.
\newblock {\em Advances in Neural Information Processing Systems}, 33:15232--15242.

\bibitem[Nguyen et~al., 2021]{nguyen2021robustifying}
Nguyen, V.~A., Zhang, F., Blanchet, J., Delage, E., and Ye, Y. (2021).
\newblock Robustifying conditional portfolio decisions via optimal transport.
\newblock {\em arXiv preprint arXiv:2103.16451}.

\bibitem[Nguyen et~al., 2020c]{nguyen2020distributionallypara}
Nguyen, V.~A., Zhang, X., Blanchet, J., and Georghiou, A. (2020c).
\newblock Distributionally robust parametric maximum likelihood estimation.
\newblock In {\em Advances in Neural Information Processing Systems}, volume~33, pages 7922--7932.

\bibitem[Norton et~al., 2017]{norton2017optimistic}
Norton, M., Takeda, A., and Mafusalov, A. (2017).
\newblock Optimistic robust optimization with applications to machine learning.
\newblock {\em arXiv preprint arXiv:1711.07511}.

\bibitem[Olea et~al., 2022]{olea2022generalization}
Olea, J. L.~M., Rush, C., Velez, A., and Wiesel, J. (2022).
\newblock On the generalization error of norm penalty linear regression models.
\newblock {\em arXiv preprint arXiv:2211.07608}.

\bibitem[Osogami, 2012]{osogami2012robustness}
Osogami, T. (2012).
\newblock Robustness and risk-sensitivity in {M}arkov decision processes.
\newblock In {\em Advances in Neural Information Processing Systems}, volume~25.

\bibitem[Owen, 2001]{owen2001empirical}
Owen, A.~B. (2001).
\newblock {\em Empirical {L}ikelihood}.
\newblock CRC press.

\bibitem[Peyr{\'e} et~al., 2019]{peyre2019computational}
Peyr{\'e}, G., Cuturi, M., et~al. (2019).
\newblock Computational optimal transport: With applications to data science.
\newblock {\em Foundations and Trends{\textregistered} in Machine Learning}, 11(5-6):355--607.

\bibitem[Prasad et~al., 2020]{prasad2018robust}
Prasad, A., Suggala, A.~S., Balakrishnan, S., and Ravikumar, P. (2020).
\newblock Robust estimation via robust gradient estimation.
\newblock {\em Journal of the Royal Statistical Society Series B: Statistical Methodology}, 82(3):601--627.

\bibitem[Raghavendra and Yau, 2020]{raghavendra2020list}
Raghavendra, P. and Yau, M. (2020).
\newblock List decodable learning via sum of squares.
\newblock In {\em Proceedings of the Fourteenth Annual ACM-SIAM Symposium on Discrete Algorithms}, pages 161--180. SIAM.

\bibitem[Rahimian and Mehrotra, 2022]{rahimian2019distributionally}
Rahimian, H. and Mehrotra, S. (2022).
\newblock Frameworks and results in distributionally robust optimization.
\newblock {\em Open Journal of Mathematical Optimization}, 3:1--85.

\bibitem[Rockafellar, 1974]{rockafellar1974conjugate}
Rockafellar, R. (1974).
\newblock {\em Conjugate Duality and Optimization}.
\newblock Society for Industrial and Applied Mathematics.

\bibitem[Rockafellar, 1985]{ROCKAFELLAR1985extension}
Rockafellar, R. (1985).
\newblock Extensions of subgradient calculus with applications to optimization.
\newblock {\em Nonlinear Analysis: Theory, Methods \& Applications}, 9(7):665--698.

\bibitem[Rockafellar, 1997]{rockafellar1997convex}
Rockafellar, R. (1997).
\newblock {\em Convex Analysis}.
\newblock Princeton Landmarks in Mathematics and Physics. Princeton University Press.

\bibitem[Rockafellar, 1963]{rockafellianthesis}
Rockafellar, R.~T. (1963).
\newblock {\em Convex Functions and Dual Extremum Problems}.
\newblock Phd thesis, University of Washington.

\bibitem[Rockafellar, 2023]{rockafellardistributional}
Rockafellar, R.~T. (2023).
\newblock Distributional robustness, stochastic divergences, and the quadrangle of risk.

\bibitem[Rothenh{\"a}usler and B{\"u}hlmann, 2023]{rothenhausler2023distributionally}
Rothenh{\"a}usler, D. and B{\"u}hlmann, P. (2023).
\newblock Distributionally robust and generalizable inference.
\newblock {\em Statistical Science}, 38(4):527--542.

\bibitem[Royset and Wets, 2022]{royset2022optimization}
Royset, J. and Wets, R. (2022).
\newblock {\em An Optimization Primer}.
\newblock Springer Series in Operations Research and Financial Engineering. Springer International Publishing.

\bibitem[Royset, 2021]{royset2021good}
Royset, J.~O. (2021).
\newblock Good and bad optimization models: Insights from {R}ockafellians.
\newblock In {\em Tutorials in Operations Research: Emerging Optimization Methods and Modeling Techniques with Applications}, pages 131--160. INFORMS.

\bibitem[Royset et~al., 2023]{royset2023rockafellian}
Royset, J.~O., Chen, L.~L., and Eckstrand, E. (2023).
\newblock Rockafellian relaxation in optimization under uncertainty: Asymptotically exact formulations.
\newblock {\em arXiv preprint arXiv:2204.04762}.

\bibitem[Ruszczy{\'n}ski and Shapiro, 2006]{ruszczynski2006optimization}
Ruszczy{\'n}ski, A. and Shapiro, A. (2006).
\newblock Optimization of risk measures.
\newblock {\em Probabilistic and Randomized Methods for Design under Uncertainty}, pages 119--157.

\bibitem[Sagawa et~al., 2020]{sagawa2019distributionally}
Sagawa, S., Koh, P.~W., Hashimoto, T.~B., and Liang, P. (2020).
\newblock Distributionally robust neural networks.
\newblock In {\em International Conference on Learning Representations}.

\bibitem[Santambrogio, 2015]{santambrogio2015optimal}
Santambrogio, F. (2015).
\newblock {\em Optimal Transport for Applied Mathematicians: Calculus of Variations, PDEs, and Modeling}, volume~87.
\newblock Birkh{\"a}user.

\bibitem[Scarf, 1958]{scarf1958minmax}
Scarf, H. (1958).
\newblock A min-max solution of an inventory problem.
\newblock In {\em Studies in the Mathematical Theory of Inventory and Production}, pages 201--209. Stanford University Press.

\bibitem[Scheffe and Tukey, 1944]{scheffe1944formula}
Scheffe, H. and Tukey, J.~W. (1944).
\newblock {A formula for sample sizes for population tolerance limits}.
\newblock {\em Annals of Mathematical Statistics}, 15(2):217.

\bibitem[Scheffe and Tukey, 1945]{scheffe1945nonpara}
Scheffe, H. and Tukey, J.~W. (1945).
\newblock {Non-parametric estimation. I. Validation of order statistics}.
\newblock {\em Annals of Mathematical Statistics}, 16(2):187 -- 192.

\bibitem[Shafieezadeh-Abadeh et~al., 2023]{shafieezadeh2023new}
Shafieezadeh-Abadeh, S., Aolaritei, L., D{\"o}rfler, F., and Kuhn, D. (2023).
\newblock New perspectives on regularization and computation in optimal transport-based distributionally robust optimization.
\newblock {\em arXiv preprint arXiv:2303.03900}.

\bibitem[Shafieezadeh-Abadeh et~al., 2019]{shafieezadeh2019regularization}
Shafieezadeh-Abadeh, S., Kuhn, D., and Esfahani, P.~M. (2019).
\newblock Regularization via mass transportation.
\newblock {\em Journal of Machine Learning Research}, 20(103):1--68.

\bibitem[Shafieezadeh~Abadeh et~al., 2018]{shafieezadeh2018wasserstein}
Shafieezadeh~Abadeh, S., Nguyen, V.~A., Kuhn, D., and Mohajerin~Esfahani, P.~M. (2018).
\newblock Wasserstein distributionally robust {K}alman filtering.
\newblock In {\em Advances in Neural Information Processing Systems}, volume~31.

\bibitem[Shapiro, 2017]{shapiro2017distributionally}
Shapiro, A. (2017).
\newblock Distributionally robust stochastic programming.
\newblock {\em SIAM Journal on Optimization}, 27(4):2258--2275.

\bibitem[Si et~al., 2021]{si2021testing}
Si, N., Murthy, K., Blanchet, J., and Nguyen, V.~A. (2021).
\newblock Testing group fairness via optimal transport projections.
\newblock In {\em International Conference on Machine Learning}, pages 9649--9659. PMLR.

\bibitem[Si et~al., 2023]{si2023distributionally}
Si, N., Zhang, F., Zhou, Z., and Blanchet, J. (2023).
\newblock Distributionally robust batch contextual bandits.
\newblock {\em Management Science}.

\bibitem[Sinha et~al., 2018]{sinha2018certifying}
Sinha, A., Namkoong, H., and Duchi, J. (2018).
\newblock Certifying some distributional robustness with principled adversarial training.
\newblock In {\em International Conference on Learning Representations}.

\bibitem[Staib and Jegelka, 2019]{staib2019distributionally}
Staib, M. and Jegelka, S. (2019).
\newblock Distributionally robust optimization and generalization in kernel methods.
\newblock In {\em Advances in Neural Information Processing Systems}, volume~32.

\bibitem[Steinhardt et~al., 2018]{steinhardt2018resilience}
Steinhardt, J., Charikar, M., and Valiant, G. (2018).
\newblock Resilience: {A} criterion for learning in the presence of arbitrary outliers.
\newblock In {\em 9th Innovations in Theoretical Computer Science Conference}, volume~94, pages 45:1--45:21.

\bibitem[Steinhardt et~al., 2017]{steinhardt2017certified}
Steinhardt, J., Koh, P.~W., and Liang, P. (2017).
\newblock Certified defenses for data poisoning attacks.
\newblock In {\em Advances in Neural Information Processing Systems}, volume~30, page 3520–3532.

\bibitem[Stone, 1977]{stone1977consistent}
Stone, C.~J. (1977).
\newblock Consistent nonparametric regression.
\newblock {\em The annals of statistics}, pages 595--620.

\bibitem[Strassen, 1965]{strassen1965existence}
Strassen, V. (1965).
\newblock The existence of probability measures with given marginals.
\newblock {\em Annals of Mathematical Statistics}, 36(2):423--439.

\bibitem[Suggala et~al., 2019]{suggala2019adaptive}
Suggala, A.~S., Bhatia, K., Ravikumar, P., and Jain, P. (2019).
\newblock Adaptive hard thresholding for near-optimal consistent robust regression.
\newblock In {\em Conference on Learning Theory}, pages 2892--2897. PMLR.

\bibitem[Sun and Zou, 2021]{sun2021data}
Sun, Z. and Zou, S. (2021).
\newblock A data-driven approach to robust hypothesis testing using kernel mmd uncertainty sets.
\newblock In {\em 2021 IEEE International Symposium on Information Theory (ISIT)}, pages 3056--3061. IEEE.

\bibitem[Székely, 1989]{szekely1989potential}
Székely, G.~J. (1989).
\newblock Potential and kinetic energy in statistics.
\newblock Lecture Notes, Budapest Institute of Technology (Technical University).

\bibitem[Taskesen et~al., 2020]{taskesen2020distributionally}
Taskesen, B., Nguyen, V.~A., Kuhn, D., and Blanchet, J. (2020).
\newblock A distributionally robust approach to fair classification.
\newblock {\em arXiv preprint arXiv:2007.09530}.

\bibitem[Taskesen et~al., 2021]{taskesen2021sequential}
Taskesen, B., Yue, M.-C., Blanchet, J., Kuhn, D., and Nguyen, V.~A. (2021).
\newblock Sequential domain adaptation by synthesizing distributionally robust experts.
\newblock In {\em International Conference on Machine Learning}, pages 10162--10172. PMLR.

\bibitem[Tibshirani, 1996]{tibshirani1996regression}
Tibshirani, R. (1996).
\newblock Regression shrinkage and selection via the lasso.
\newblock {\em Journal of the Royal Statistical Society Series B: Statistical Methodology}, 58(1):267--288.

\bibitem[Tukey, 1960]{tukey1959survey}
Tukey, J.~W. (1960).
\newblock A survey of sampling from contaminated distributions.
\newblock {\em Contributions to Probability and Statistics: Essays in Honor of Harold Hotelling}, pages 448--485.

\bibitem[Tukey, 1962]{Tukey1962future}
Tukey, J.~W. (1962).
\newblock The future of data analysis.
\newblock {\em Annals of Mathematical Statistics}, 33(1):1--67.

\bibitem[Tukey, 1975]{Tukey1975Mathematics}
Tukey, J.~W. (1975).
\newblock Mathematics and the picturing of data.
\newblock In {\em Proceedings of the International Congress of Mathematicians}, volume~2, page 523–531.

\bibitem[Vaart, 1998]{vaart1998asymptotic}
Vaart, A. W. v.~d. (1998).
\newblock {\em Asymptotic Statistics}.
\newblock Cambridge Series in Statistical and Probabilistic Mathematics. Cambridge University Press.

\bibitem[Van~Parys et~al., 2021]{van2021data}
Van~Parys, B.~P., Esfahani, P.~M., and Kuhn, D. (2021).
\newblock From data to decisions: Distributionally robust optimization is optimal.
\newblock {\em Management Science}, 67(6):3387--3402.

\bibitem[Villani et~al., 2009]{villani2009optimal}
Villani, C. et~al. (2009).
\newblock {\em Optimal Transport: Old and New}, volume 338.
\newblock Springer.

\bibitem[Wang et~al., 2021]{wang2021sinkhorn}
Wang, J., Gao, R., and Xie, Y. (2021).
\newblock Sinkhorn distributionally robust optimization.
\newblock {\em arXiv preprint arXiv:2109.11926}.

\bibitem[Wang et~al., 2023]{wang2023foundation}
Wang, S., Si, N., Blanchet, J., and Zhou, Z. (2023).
\newblock On the foundation of distributionally robust reinforcement learning.
\newblock {\em arXiv preprint arXiv:2311.09018}.

\bibitem[Watson and Holmes, 2016]{watson2016approximate}
Watson, J. and Holmes, C. (2016).
\newblock Approximate models and robust decisions.
\newblock {\em Statistical Science}, 31(4):465--489.

\bibitem[Weiss et~al., 2016]{weiss2016survey}
Weiss, K., Khoshgoftaar, T.~M., and Wang, D. (2016).
\newblock A survey of transfer learning.
\newblock {\em Journal of Big data}, 3(1):1--40.

\bibitem[Wilson and Cook, 2020]{wilson2020survey}
Wilson, G. and Cook, D.~J. (2020).
\newblock A survey of unsupervised deep domain adaptation.
\newblock {\em ACM Transactions on Intelligent Systems and Technology (TIST)}, 11(5):1--46.

\bibitem[Wu et~al., 2020]{wu2020minimax}
Wu, K., Ding, G.~W., Huang, R., and Yu, Y. (2020).
\newblock On minimax optimality of gans for robust mean estimation.
\newblock In {\em International Conference on Artificial Intelligence and Statistics}, volume 108, pages 4541--4551. PMLR.

\bibitem[Xu and Mannor, 2010]{xu2010distributionally}
Xu, H. and Mannor, S. (2010).
\newblock Distributionally robust {M}arkov decision processes.
\newblock In {\em Advances in Neural Information Processing Systems}, volume~23.

\bibitem[Zalinescu, 2002]{zalinescu2002convex}
Zalinescu, C. (2002).
\newblock {\em Convex Analysis in General Vector Spaces}.
\newblock G - Reference, Information and Interdisciplinary Subjects Series. World Scientific.

\bibitem[Zhang et~al., 2022a]{zhang2022class}
Zhang, X., Blanchet, J., Ghosh, S., and Squillante, M.~S. (2022a).
\newblock A class of geometric structures in transfer learning: Minimax bounds and optimality.
\newblock In {\em International Conference on Artificial Intelligence and Statistics}, pages 3794--3820. PMLR.

\bibitem[Zhang et~al., 2022b]{zhang2022wasserstein}
Zhang, X., Blanchet, J., Marzouk, Y., Nguyen, V.~A., and Wang, S. (2022b).
\newblock Distributionally robust {G}aussian process regression and {B}ayesian inverse problems.
\newblock {\em arXiv preprint arXiv:2205.13111}.

\bibitem[Zhou et~al., 2021]{zhou2021finite}
Zhou, Z., Zhou, Z., Bai, Q., Qiu, L., Blanchet, J., and Glynn, P. (2021).
\newblock Finite-sample regret bound for distributionally robust offline tabular reinforcement learning.
\newblock In {\em International Conference on Artificial Intelligence and Statistics}, pages 3331--3339. PMLR.

\bibitem[Zhu et~al., 2022]{zhu2022generalized}
Zhu, B., Jiao, J., and Steinhardt, J. (2022).
\newblock {Generalized resilience and robust statistics}.
\newblock {\em Annals of Statistics}, 50(4):2256 -- 2283.

\bibitem[Zhu et~al., 2021]{zhu2021kernel}
Zhu, J.-J., Jitkrittum, W., Diehl, M., and Sch{\"o}lkopf, B. (2021).
\newblock Kernel distributionally robust optimization: Generalized duality theorem and stochastic approximation.
\newblock In {\em International Conference on Artificial Intelligence and Statistics}, pages 280--288. PMLR.

\bibitem[Zorzi, 2016]{zorzi2016robust}
Zorzi, M. (2016).
\newblock Robust {K}alman filtering under model perturbations.
\newblock {\em IEEE Transactions on Automatic Control}, 62(6):2902--2907.

\end{thebibliography}
\bibliographystyle{apalike}

\end{document}